\documentclass[10pt,authoryear]{article}
\usepackage{appendix}
\usepackage{setspace}
\usepackage[font=small,  margin=2cm]{caption}
\usepackage[tbtags]{amsmath}
\usepackage{amsthm,amssymb,amsfonts}
\usepackage{natbib}
\usepackage{multirow}
\usepackage{lscape}
\usepackage{subfigure}
\usepackage{makecell}
\usepackage{exscale}
\usepackage{booktabs}
\usepackage{array}
\usepackage{fullpage}
\usepackage{url}
\usepackage{algorithm}
\usepackage{algpseudocode}
\usepackage{bm}
\usepackage{smile}
\usepackage{mathtools}
\usepackage{wrapfig}
\usepackage{lipsum}
\usepackage{threeparttable}
\usepackage{mathrsfs}
\usepackage{relsize}
\usepackage{dsfont}
\usepackage{multirow}
\usepackage{apalike}
\usepackage{subfigure}
\usepackage[usenames,dvipsnames,svgnames,table]{xcolor}
\usepackage[colorlinks=true,linkcolor=blue,urlcolor=blue,citecolor=blue]{hyperref}
\usepackage{stfloats}
\usepackage{float}
\makeatletter
\newcommand\fcaption{\def\@captype{figure}\caption}
\makeatother

\numberwithin{equation}{section}
\numberwithin{theorem}{section}
\numberwithin{corollary}{section}
\numberwithin{definition}{section}

\begin{document}
	\title{\LARGE Transfer Learning in High-dimensional Semi-parametric Graphical Models with Application to Brain Connectivity Analysis}
	
	\author{Yong He\thanks{ Institute for Financial Studies, Shandong University, Jinan, China; Email:{\tt heyong@sdu.edu.cn}.},~~Qiushi Li\thanks{Institute for Financial Studies, Shandong University, Jinan, China; Email:{\tt lqs\_chelsea@126.com}.},~~Qinqin Hu\thanks{ School of Mathematics and Statistics, Shandong University, Weihai, Weihai, China; Email:{\tt qqhu@sdu.edu.cn}. Corresponding Author},~~Lei Liu\thanks{Division of Biostatistics, Washington University in St. Louis, U.S.A.;  Email:{\tt lei.liu@wustl.edu}}.}	
	
\date{}	
\maketitle
Transfer learning has drawn growing attention with the target of improving statistical efficiency of one study (dataset) by digging information from similar and related auxiliary studies (datasets). In the article, we consider transfer learning problem in estimating undirected semi-parametric graphical model.
We propose an algorithm called  Trans-Copula-CLIME for estimating undirected graphical model while digging information from similar auxiliary studies, characterizing the similarity between the target graph and each auxiliary graph by the sparsity of a divergence matrix. The proposed method relaxes the restrictive assumption that data follows a Gaussian distribution, which deviates from reality for the fMRI dataset  related to  Attention
Deficit Hyperactivity Disorder (ADHD) considered here. Nonparametric rank-based correlation coefficient estimators are
utilized in the Trans-Copula-CLIME procedure to achieve robustness against normality. We establish the convergence rate of the  Trans-Copula-CLIME estimator  under some mild conditions, which demonstrates that when the similarity between the auxiliary studies and the target study is sufficiently high and the number of informative auxiliary samples is sufficiently large, then the  Trans-Copula-CLIME estimator shows great advantage over the existing non-transfer-learning ones. Simulation studies also show that  Trans-Copula-CLIME estimator has better performance especially when  data are  not from Gaussian distribution. At last, the proposed method is applied to infer functional brain connectivity pattern for ADHD patients in the target  Beijing site by leveraging the fMRI datasets from New York site.
\vspace{2em}


\section{Introduction}
Brain connectivity analysis (BCA) has nowadays been at the foreground of neuroscience research with the aid of modern imaging technology such as functional Magnetic Resonance Imaging (fMRI), which
reveals the synchronization of brain systems through correlations in the neurophysiological measurement of brain
activities. The main tool for BCA is graphical model, represented by $\mathcal{G}=(V,E)$, where $V=\{1,\ldots,p\}$ is the set of vertices and $E$ the set of edges in $V\times V$.  In BCA, a node in $V$ represents a brain region and an edge in $E$ represents functional connectivity between  the  brain regions at the end of the edge.
  One of the most well-known graphical model is Gaussian Graphical Model (GGM), which captures partial dependence among random variables. Assume $X=(X_1,\ldots,X_p)^\top\sim N(\mu,\Sigma)$,  a pair $(i,j)$ is contained in the edge set $E$ if and only if $X_i$ is conditionally dependent of $X_j$, given all remaining variables $X_{V \setminus \{i,j\}}$ and  the absence of the pair $(i,j)$ in $E$ indicates  $X_i$ and  $X_j$ are conditionally independent.  To estimate the GGM, it is equivalent to  recover the support of the precision matrix (the inverse of the covariance matrix $\Sigma$), i.e. denote $\Omega=\Sigma^{-1}=(\Omega_{i,j})$, then $X_i\perp X_j \mid X_{V \setminus \{i,j\}}$ if and only if  $\Omega_{i,j}=0$ \citep{lauritzen96}.
  For BCA, it's often the case that the number of brain regions (dimension) $p$ is much larger than the sample size $n$, typically referred to as  high-dimensional data. Statistical inference on GGM  becomes challenging in the high-dimensional case and the past decades witness significant literature on inferring GGM, including penalized methods, see for example,  neighborhood selection method by \cite{meinshausen2006high}, graphical Lasso by \cite{friedman2008sparse,yuan2007model}, the lasso penalized D-trace loss by \cite{zhang2014sparse},  and constrained $\ell_1$  minimization approach (CLIME) by \cite{cai2011constrained}.

Modern massive and diverse imaging data sets  pose great challenge for BCA study and it is of significant importance to integrate different data sets to make  statistical inference more accurate. In addition,  BCA with the aid of graphical model based on a single study typically bears large uncertainty and low power in detecting significant brain connectivity
 in the corresponding brain network. One promising solution is the growing popular machine learning technique called transfer learning \citep{Torrey2010Transfer}.
Given a target GGM estimation problem for BCA, transfer learning  aims at transferring the knowledge from different but related imaging samples to improve the statistical
inference of the target GGM.  Transfer learning has been applied to problems in many scientific fields, including natural language processing, image classification, supervised regression, management science and sensor-based location estimation \citep{zhuang2011exploiting,raina2007self,wang2010indoor,li2020transferHDLR,bastani2021predicting}.  For GGM estimation, \cite{li2020transfer} fully exploits the information of  samples from closely related studies to estimate the graph structure of  the target study and propose an algorithm called Trans-CLIME.
Transfer learning in GGM is  different from the multi-task learning, where the goal is
to simultaneously estimate multiple graphs \citep{guo2011joint,danaher2014joint,cai38joint}.

The literature on GGM, inevitably rely on the joint normality of the variables such that penalized methods
based on least squares regression, or likelihood generally, can be applied. However, the normality assumption is simply an idealization of the complex random world.  The application that motivated the current work is the integration of the fMRI dataset related with ADHD in different sites  from the ADHD-200 Global Competition,
for understanding the  potential pathogenic mechanism of the mental disease via graphical model. These datasets are
 high-dimensional with relatively small sample sizes and deviate from the normality assumption (see Q-Q plots in Figure \ref{fig:5} in Section \ref{sec:realexa}). \cite{liu2009nonparanormal} relax the Gaussian assumption to a semiparametric Gaussian copula (Nonparanormal) assumption in graphical modelling. Instead of assuming that the original random vector $X=(X_1,X_2,\cdots,X_p)^\top$ follows a Gaussian distribution, they assume that the transformed random vector $f(X)=(f_1(X_1),f_2(X_2),\ldots,f_p(X_p))^\top$ is  multivariate normal distributed, where $f=\{f_1,\cdots,f_p\}$ is a set of monotone univariate functions.  \cite{liu2012high} and \cite{xue2012regularized} propose regularized rank-based estimation method to infer the semiparametric Gaussian copula graphical model, which achieve the optimal parametric rates of convergence for both graph recovery and precision matrix estimation.   When studying the brain connectivity structure of ADHD from a specific site, it
is helpful to incorporate  auxiliary information from other sites to further enhance the graphical
learning accuracy. Also in view of the non-normality of fMRI data, we are motivated to consider transfer learning in
high-dimensional nonparanormal graphical model.

In this article, we propose a method called  Trans-Copula-CLIME, which not only relaxes the normality assumption in GGM but also can achieve the goal of transfer learning by integrating datasets from auxiliary studies. We establish the convergence rate of  Trans-Copula-CLIME estimator and demonstrate that if the similarity between the auxiliary studies and the target study is sufficiently strong and the number of informative auxiliary samples is sufficiently large, then the  Trans-Copula-CLIME estimator achieves a faster convergence rate than the nonparametric estimator in \cite{liu2012high} from a single study under some mild conditions, in terms of precision matrix estimation. Thorough simulation studies  show that  Trans-Copula-CLIME estimator has comparable performance with Trans-CLIME by \cite{li2020transfer} when Gaussian assumption is satisfied and  has better empirical performance  when the distribution of data deviates from Gaussian. Therefore, when we have a sufficient number of auxiliary samples similar to the target study and are not sure whether the data satisfies the Gaussian assumption, we can always resort to the  Trans-Copula-CLIME to estimate the graphical structure of the target study and this estimator performs no worse than the existing ones. As far as we know, this is the first work on transfer learning for semi-parametric undirected  graphical modelling with statistical guarantee.

The rest of the paper proceeds as follows. In Section 2, we briefly introduce some preliminary results on transfer learning in Gaussian graphic model and the Gaussian copula/ Nonparanormal distribution.  In Section 3, we present the Trans-Copula-CLIME algorithm. In Section 4, we establish the convergence rate for the proposed Trans-Copula-CLIME estimator. In Section 5, we provide a thorough numerical simulation study. A real dataset of ADHD is presented in Section 6. Section 7 concludes the paper. The proofs of the main theorems are given in the Appendix.

\section{Preliminaries}
In this section we introduce the Gaussian copula distribution and some preliminary results on transfer learning of Gaussian Graphical Model. Prior to this, we first introduce the notations adopted throughout the paper. Denote $1_p$  as a $p$-dimensional vector with all elements equal to 1. Denote $I(\cdot)$ as the indicator function. For any vector $x=(x_1,\ldots,x_p)^\top \in \RR^p$, let $\|x\|_1=\sum_{i=1}^p|x_i|$, $ \|x\|_2=(\sum_{i=1}^px_i^2)^{1/2}$, $\|x\|_{\infty}=\max_{i\leqslant p}|x_i|$. For a matrix $A \in \mathbb{R}^{p \times p}$, let $\mathrm{A}_{ij}$ be the $ij$ entry of $A$, $A^\top$ be the transpose of $A$. Let $A_j$ denote the $j$-th column of $A$. For any fixed $j\le p$, let $\|A_j\|_2$ denote the column-wise $\ell_2$-norm of $A$. Let $\|A\|_{\infty,2}=\max_{j\le p}\|A_j\|_2$, $\|A\|_{1}=\max_{j\le p}\|A_j\|_1$, $\|A\|_{\max}=\max_{i,j\le p}|A_{ij}|$, and $\|A\|_1=\sum_{j=1}^{p}\|A_j\|_1$. Let $\|A\|$ be the spectral norm of matrix $A$ and $\|A\|_F$ be the Frobenius norm of $A$. Denote $\Lambda_{\max}(A)$ and $\Lambda_{\min}(A)$ as the largest and smallest eigenvalues of a nonnegative definitive matrix $A$, respectively. We use $c_0, c_1, \dots$ and $M,C_0,C_1,\dots$ as generic constants which can be different at different places.
\subsection{Gaussian Copula distribution}\label{sec:2.2}
The Gaussian Graphic models(GGM) have been widely used in biology, finance and other fields. However, GGM assume  random vectors follow  multivariate normal distribution, which is a relatively restrictive in real application. \cite{liu2009nonparanormal} relax the Gaussian assumption in graphical modelling to the semi-parametric Gaussian copula assumption, also known as Nonparanormal distribution in the literature. Instead of assuming that the original random vector $X=(X_1,X_2,\ldots,X_p)^\top$ follows the Gaussian distribution, they assume that the transformed random vector $f(X)=(f_1(X_1),f_2(X_2),\ldots,f_p(X_p))^\top$ follows multivariate normal distribution. More precisely, we have the following definition.
\begin{definition}
	(Gaussian Copula distribution) Let $f=\{f_1,\cdots,f_p\}$ be a set of monotone univariate functions and let $\Sigma\in \mathbb{R}^{p \times p}$ be a positive-definite correlation matrix with $\text{diag}(\Sigma)=1_p$. We say a $p$-dimensional random variable $X=(X_1,X_2,\ldots,X_p)^\top$ follows Gaussian Copula distribution, if $f(X)=(f_1(X_1),f_2(X_2),\ldots,f_p(X_p))^\top\sim N_p(0,\Sigma)$, denoted as $X\sim \text{NPN}_p(f,\Sigma)$.
\end{definition}
\cite{liu2009nonparanormal} show that the precision matrix $\Omega= \Sigma^{-1}$ captures the  conditional dependence structure of $X$, that is, $X_j$ and $X_k$ are independent given the rest of the variables in $X$ if and only if $\Omega_{jk}=0$. Therefore, to estimate the graph for the Gaussian copula family, it suffices to recover the support set  of $\Omega$.
\subsection{Transfer learning in Gaussian Graphical Model}\label{sec:2.3}
Supposed that $i.i.d.$ observations $x_1,\dots,x_n \in \mathbb{R}^p$ are generated from the target distribution $N(0,\Sigma)$, and $K$ auxiliary studies $x_i^{(k)}\in \mathbb{R}^p$ are independently generated from $N(0,\Sigma^{(k)}),i=1,\dots,n_k, \ k=1,\dots,K$. \cite{li2020transfer} define the $k$-th divergence matrix $\Delta^{(k)}=\Omega\Sigma^{(k)}-I_p$ to motivate the similarity measure between the $k$-th auxiliary study and the target one and define the auxiliary studies that are within a certain range of similarity with the target study as the informative set $\mathcal{A}$. They further define the weighted average of the covariance and divergence matrices $\Sigma^{\mathcal{A}}=\sum\limits_{k\in\mathcal{A}}\alpha_k\Sigma^{(k)}$ and
$\Delta^{\mathcal{A}}=\sum\limits_{k\in\mathcal{A}}\alpha_k\Delta^{(k)}$, where $\alpha_k=n_k/n_{\mathcal{A}}$ for $n_{\mathcal{A}}=\sum\limits_{k\in\mathcal{A}}n_k$. Using the divergence matrix, they obtain the following equation:
\begin{equation}\label{eq1}
	\Sigma\Delta^{\mathcal{A}}-(\Sigma^{\mathcal{A}}-\Sigma)=0,
\end{equation}
\begin{equation}\label{eq2}
	\Sigma^{\mathcal{A}}\Omega-(\Delta^{\mathcal{A}})^{\top}-I_p=0.
\end{equation}
They first estimate $\Delta^{\mathcal{A}}$ via equation (\ref{eq1}) under sparsity assumption, by a similar idea as CLIME.  Then by plugging $\widehat{\Delta}^{\mathcal{A}}$ into equation (\ref{eq2}), they estimate the target sparse $\Omega$. Under some mild conditions, they prove that if the similarity between target study and auxiliary studies is sufficiently strong and the sample size of auxiliary studies is much larger than the target sample size, then estimator obtained by Trans-CLIME has faster convergence rate than the CLIME estimator from a single study.

\section{Methodology}
In this section, we relax Trans-CLIME's assumption of normality to Gaussian copula distribution and propose an algorithm called Trans-Copula-CLIME.

\subsection{Rank-based estimator}\label{Rbe}
Let $\Sigma,\Sigma^{(1)},\dots,\Sigma^{(K)}\in\mathbb{R}$ be positive-definite correlation	matrices and denote the precision matrix $\Omega=\Sigma^{-1}$. Suppose that  $x_1,\dots,x_n \in \mathbb{R}^p$ are $i.i.d.$ observations generated from the target distribution $\text{NPN}(f,\Sigma)$, and $K$ auxiliary studies $x_i^{(k)}\in \mathbb{R}^p$ are independently generated from $\text{NPN}(f^{(k)},\Sigma^{(k)}),i=1,\dots,n_k,k=1,\dots,K$. Instead of firstly estimating the marginal transformation functions and then calculating correlation matrix of the transformed data, we exploit and Kendall's tau statistics to directly estimate the correlation matrices.

The population version of Kendall's tau correlation between the random variable $X_i$ and $X_j$ is given by $\tau_{ij} := \text{Corr}(\sign(X_i-\tilde{X}_i),\sign(X_j-\tilde{X}_j)$, where $\widetilde{X}_i$ and $\widetilde{X}_j$ are two independent copies of $X_i$ and $X_j$.
Let $x_1, \dots , x_n \in \mathbb{R}^p$ be $n$ data points and  $x_s=(x_{s1}, \ldots,x_{sp})^\top, s=1\ldots,n$. Then the Kendall's tau correlation between the empirical realizations of random variable $X_i$ and $X_j$ is defined as
\[\widehat{\tau}_{ij}=\frac{2}{n(n-1)}\sum_{1\le m<m^{\prime}\le n} \sign\big((x_{mi}-x_{m^{\prime}i})(x_{mj}-x_{m^{\prime}j})\big).\]

In fact, there is a relationship between Kendall's tau and Pearson correlation coefficient for multivariate normal distribution. That is $\Sigma_{ij}=\sin(\frac{\pi}{2}\tau_{ij})$, to which one may refer \cite{kendall1948rank}. Note that Kendall's tau correlation is invariant under monotone transformation. Therefore, we  define the following estimator $\widehat{S}=[\widehat{S}_{ij}]$ for the unknown correlation matrix $\Sigma$ of the target study:
$\widehat{S}_{ij} = \sin(\frac{\pi}{2}\widehat{\tau}_{ij})I(i=j)+I(i\neq j) $
Also, we can estimate the correlation matrix $\Sigma^{(k)}$ of the $k$-th auxiliary study in a similar way and denote the corresponding estimators as $\widehat{S}^{(k)},k=1,2,\dots,K$.
\subsection{Trans-Copula-CLIME algorithm}
In this subsection, we explain how to exploit the estimated correlation matrices $\widehat{S}$ and $\widehat{S}^{(k)},k=1,2\dots,K$ to estimate the sparse precision matrix of the target study.

We similarly define the $k$-th divergence matrix $\Delta^{(k)}=\Omega\Sigma^{(k)}-I_p$, the weighted average of the correlation and divergence matrices $\Sigma^{\mathcal{A}}=\sum\limits_{k\in\mathcal{A}}\alpha_k\Sigma^{(k)}$ and
$\Delta^{\mathcal{A}}=\sum\limits_{k\in\mathcal{A}}\alpha_k\Delta^{(k)}$, where $\alpha_k=n_k/n_{\mathcal{A}}$ for $n_{\mathcal{A}}=\sum\limits_{k\in\mathcal{A}}n_k$. We find the similar equation is also true for the correlation matrix:
\begin{equation}\label{eq3}
	\Sigma\Delta^{\mathcal{A}}-(\Sigma^{\mathcal{A}}-\Sigma)=0,
\end{equation}
and
\begin{equation}\label{eq4}
	\Sigma^{\mathcal{A}}\Omega-(\Delta^{\mathcal{A}})^{\top}-I_p=0.
\end{equation}

Therefore, we  first estimate $\Delta^{\mathcal{A}}$ via equation (\ref{eq3}) under sparsity assumption, then estimate the target sparse $\Omega$ by plugging $\widehat{\Delta}^{\mathcal{A}}$ into equation (\ref{eq4}). $\widehat{S}$ and $\widehat{S}^{(k)}$, as defined in Section \ref{Rbe}, are the estimators of the correlation matrix based on Kendall's tau of the target study and auxiliary studies respectively. Let $\widehat{S}^{\mathcal{A}}=\sum\limits_{k\in\mathcal{A}}\alpha_k\widehat{S}^{(k)}$ denote the sample rank-based correlation matrix from the informative auxiliary samples. In the following, we introduce the specific steps of our proposed transfer learning algorithm, viz., Trans-Copula-CLIME.

\textbf{Step 1}. Compute the single-study Copula CLIME estimator
\begin{equation}\label{opcl}
		\widehat{\Omega}^{\text{(CL)}}=\underset{\Omega\in\mathbb{R}^{p\times p}}{\arg\min}\|\Omega\|_1, \ \
		\text {subject to } \ \ \|\widehat{S}\Omega-I_p\|_{\max}\leqslant\lambda_{\text{CL}},
\end{equation}
which are used in step 2.

\textbf{Step 2}. Compute
\begin{equation}\label{opS1}
		\widehat{\Delta}^{(0)}=\underset{\Delta\in\mathbb{R}^{p\times p}}{\arg \min}\|\Delta\|_1, \ \
		\text {subject to }\ \ \|\widehat{S}\Delta-(\widehat{S}^{\mathcal{A}}-\widehat{S})\|_{\max}\leqslant\lambda_{\Delta}.
\end{equation}
As the obtained $\widehat{\Delta}^{(0)}$ is not necessarily row-wise sparse, we refine it as follows:
\begin{equation}\label{opS1db}
		\widehat{\Delta}^{\mathcal{A}}=\underset{\Delta\in\mathbb{R}^{p\times p}}{\arg\min}\|\Delta\|_1, \ \
		\text {subject to }\ \ \|\Delta-\widehat{\Delta}^{(0)}-\widehat{\Omega}^{\text{(CL)}}(\widehat{S}^{\mathcal{A}}-\widehat{S}-\widehat{S}\widehat{\Delta}^{(0)})\|_{\max}\leqslant 2\lambda_{\Delta}.
\end{equation}

\textbf{Step 3}. For $\widehat{\Delta}^{\mathcal{A}}$ defined in (\ref{opS1db}), compute
\begin{equation}\label{opS2}
		\widehat{\Omega}=\underset{\Omega\in\mathbb{R}^{p\times p}}{\arg\min}\|\Omega\|_1, \ \
		\text {subject to }\ \ \|\widehat{S}^{\mathcal{A}}\Omega-(\widehat{\Delta}^{\mathcal{A}}+I_p)^{\top}\|_{\max}\leqslant\lambda_{\Omega}.
\end{equation}

Computationally, all the optimizations in the above three steps can be decomposed into $p$ minimization problems, which makes the computation scalable to large datasets.
\begin{remark}\label{agg}
	If the similarity between the target study and the auxiliary studies is very low or the sample size of the informative auxiliary studies is much smaller than the  sample size of the target study, the output estimator of this algorithm may be not as good as the single-study Copula CLIME estimator obtained in step 1. One solution is to perform an aggregation step, similarly as \cite{li2020transfer}. To this end, we split data from the target study into two folds $\mathcal{I}$ and $\mathcal{I}^c$. Let $\widetilde{S}$ be the estimator of the correlation matrix $\Sigma$ using data from $\mathcal{I}$, while $\widetilde{S}^c$ be the estimator using data from $\mathcal{I}^c$. We use $\widetilde{S}$ in the above Steps 1-3 to calculate $\widehat{\Omega}^{\text{(CL)}}$ and $\widehat{\Omega}$ and then for $j=1,\dots,p$, compute
	\[\widehat{W}(j)=\begin{pmatrix}
		(\widehat{\Omega}_j^{\text{(CL)}})^{\top}\widetilde{S}^c\widehat{\Omega}_j^{\text{(CL)}} & (\widehat{\Omega}_j^{\text{(CL)}})^{\top}\widetilde{S}^c\widehat{\Omega}_j \\
		(\widehat{\Omega}_j^{\text{(CL)}})^{\top}\widetilde{S}^c\widehat{\Omega}_j & \widehat{\Omega}_j^{\top}\widetilde{S}^c\widehat{\Omega}_j
	\end{pmatrix},\ \hat{v}_j=\{\widehat{W}(j)\}^{-1}\begin{pmatrix}
		\widehat{\Omega}_{jj}^{\text{(CL)}} \\ \widehat{\Omega}_{jj}
	\end{pmatrix}\in\mathbb{R}^2.\]
	For $j=1,\dots,p$, let $\widetilde{\Omega}_j=(\widehat{\Omega}_j^{(\text{CL})},\widehat{\Omega}_j)\hat{v}_j$, which is the final estimator after aggregation step. Although the theoretical property of this estimator is still not clear, it has good performance according to our simulation study.
\end{remark}

\section{Theoretical results}
In this section, we analyze the statistical property of the Trans-Copula-CLIME estimator. We assume the following conditions hold in our theoretical analysis.

\vspace{0.5em}

 \textbf{Assumption A}   Assume that $x_i \in \mathbb{R}^p$ are $i.i.d.$ distributed as $\text{NPN}_p(f,\Sigma)$ for $i=1,\dots n$. For each $k \in \mathcal{A}$, $x_i^{(k)}$ are $i.i.d.$ distributed as $\text{NPN}_p(f^{(k)},\Sigma^{(k)})$ for $i=1,\dots, n_k$.

 \vspace{0.5em}

 \textbf{Assumption B}   Assume that $1/C\leqslant\Lambda_{\min}(\Sigma)\leqslant\Lambda_{\max}(\Sigma)\leqslant C$, $1/C\leqslant\min_{k\in\mathcal{A}}\Lambda_{\min}(\Sigma^{(k)})\leqslant \max_{k\in\mathcal{A}}$ $\Lambda_{\max}(\Sigma^{(k)})\leqslant C$, $\|\Omega\|_1\leqslant M$ and $\|\Omega^{\mathcal{A}}\|_1\leqslant M$, where $\Omega^{\mathcal{A}}=(\Sigma^{\mathcal{A}})^{-1}$.

\vspace{0.5em}

The parameter space we consider is
\[\mathbb{G}(s,h)=\bigg\{(\Omega,\Omega^{(1)},\dots,\Omega^{(K)}):\max\limits_{1\leqslant j\leqslant p}\|\Omega_j\|_0\leqslant s,\ \max\limits_{k \in \mathcal{A}}\mathcal{D}(\Omega,\Omega^{(k)})\leqslant h\bigg\},\]
where $\mathcal{D}(\Omega,\Omega^{(k)})=\max\limits_{1\leqslant j\leqslant p}\|\Delta_{j,.}^{(k)}\|_1+\max\limits_{1\leqslant j\leqslant p}\|\Delta_{.,j}^{(k)}\|_1$ is a measure of difference between $\Omega$ and $\Omega^{(k)}$.


The following theorem gives the convergence rate for the  Trans-Copula-CLIME estimator under the aforementioned conditions.
\begin{theorem}\label{t1}
	(Convergence rate of  Trans-Copula-CLIME). Suppose that $(\Omega,\Omega^{(1)},\dots,\Omega^{(K)})\in\mathbb{G}(s,h)$ and Assumptions A and B hold. Let $\delta_n=\sqrt{\log p/n}\wedge h$. Let the  Trans-Copula-CLIME estimator $\widehat{\Omega}$ be computed with
	\[\lambda_{\Delta}=c_1\sqrt{\frac{\log p}{n}}\text{,}\quad \text{and}\ \lambda_{\Omega}=c_2\sqrt{\frac{\log p}{n_{\mathcal{A}}}}\text{,}\]
	where $c_1$ and $c_2$ are large enough constants. If $s^2\log p=o(n_{\mathcal{A}})$, $h \lesssim s \sqrt{\log p / n}\leqslant c_3$, $h\delta_n=O(\lambda_{\Omega})$ and $n_{\mathcal{A}} \gtrsim n$, then we have
	\[\|\widehat{\Omega}_j-\Omega_j\|_2^2\vee\frac{1}{p}\|\widehat{\Omega}-\Omega\|_F^2=O_P\bigg(\frac{s\log p}{n_{\mathcal{A}}}+h\delta_n\bigg),\] for any fixed $1\le j\le p$.
\end{theorem}

Theorem \ref{t1} demonstrates that the convergence rate of  Trans-Copula-CLIME estimator relies on $n_\mathcal{A}$ and $h$, which respectively represent the number of informative auxiliary samples and the similarity between informative auxiliary studies and the target study. The larger the number of informative auxiliary samples and the stronger the similarity between the auxiliary studies and the target study, the faster the convergence rate of  the Trans-Copula-CLIME estimator will be.

To illustrate the advantage of the transfer learning algorithm, we compare the  results in Theorem \ref{t1} with the convergence rate of Copula CLIME estimator $\widehat{\Omega}^{\text{(CL)}}$ defined in (\ref{opcl}) in a single study setting.

\begin{theorem}\label{t2}
	Suppose that $\Omega\in\big\{\Omega:\max\limits_{1\leqslant j\leqslant p}\|\Omega_j\|_0\leqslant s\big\}$ and Assumptions A and B hold. Let $\lambda_{\text{CL}}=c_1\sqrt{\log p/n}$ with large enough $c_1$. If $s^2\log p=o(n)$, then for the Copula CLIME estimator $\widehat{\Omega}^{\text{(CL)}}$,
	\[\|\widehat{\Omega}_j^{\text{(CL)}}-\Omega_j\|_2^2\vee\frac{1}{p}\|\widehat{\Omega}^{\text{(CL)}}-\Omega\|_F^2=O_P\bigg(\frac{s\log p}{n}\bigg)\]
	for any fixed $1\le j\le p$.
\end{theorem}

We see that the convergence rate of  Trans-Copula-CLIME estimator  is no slower than the Copula CLIME estimator if the conditions of the Theorem \ref{t1} are satisfied, in terms of both column-wise $\ell_2$-norm and Frobenius norm. Furthermore, when $n_{\mathcal{A}}\gg n$ and $h\delta_n\ll s\log p/n$, the convergence rate of  Trans-Copula-CLIME estimator is faster. Hence, if we have a sufficient number of auxiliary samples similar to the target study and are not sure whether the data satisfies the Gaussian assumption, we can always choose the  Trans-Copula-CLIME to estimate the precision matrix of the target study and this estimator performs no worse than the existing ones.

\section{Simulation Study}
In this section, we conduct thorough simulation studies to compare the Trans-Copula-CLIME estimator with the existing methods. In particular, we consider the following methods for comparison: (i) C: the original CLIME estimator from \cite{cai2011constrained}, which only uses the data from the target study; (ii) TC: the Trans-CLIME estimator from \cite{li2020transfer}; (iii) PC: the Pooled Trans-CLIME estimator, which uses both informative and non-informative auxiliary data; (iv) CC: the Copula CLIME estimator from (\ref{opcl}); (v) CTC: the  Trans-Copula-CLIME estimator proposed in this paper; (vi) CPC: the Copula Pooled Trans-CLIME estimator, which also uses both informative and non-informative auxiliary data. The first three methods are considered for comparison in \cite{li2020transfer}, which relies on the Gaussian assumption. The last three methods are based on Gaussian Copula distribution. PC and CPC  are used to study the robustness of the estimator. The four methods in which transfer learning are involved all consider an aggregation step mentioned in Remark \ref{agg}.
\subsection{Numerical Setup}
We set $p=100,n=200,n_k=200,K=5$ for $k=1,2,\dots,K$ and consider two types of precision matrix $\Omega$ which adopted in \cite{li2020transfer}:

\vspace{0.5em}

(i) Banded matrix with bandwidth 8. For $1 \leqslant i,j \leqslant p,\Omega_{ij}=2\times0.6^{|i-j|}{I}(|i-j|\leqslant 7)$.

\vspace{0.5em}

(ii) Block diagonal matrix with block size 4, where each block is Toeplitz (1.2, 0.9, 0.6, 0.3).

\vspace{0.5em}

Let $\Sigma=\Omega^{-1}$ and to obtain the correlation matrix, we simply rescale $\Sigma$ so that all its diagonal elements are 1. The inverse of the rescaled $\Sigma$ is used as the final precision matrix. We use the following steps to obtain $\Omega^{(k)}$. For $k \in \mathcal{A}$, we first generate $\Delta^{(k)}$. The $(i,j)$-entry of $\Delta^{(k)}$ is zero with probability 0.9 and is nonzero with probability 0.1. If an entry is nonzero, it is randomly generated from uniform distribution $U\left[-r/p,r/p\right]$ for $r\in\{10,20,30\}$, where $r$ is a parameter that controls the similarity level between target study and auxiliary studies. Then by the defintion of $\Delta^{(k)}:\Delta^{(k)}=\Omega\Sigma^{(k)}-I_p$, we can obtain $\Sigma^{(k)}=\Sigma(\Delta^{(k)}+I_p)$. We symmetrize $\Sigma^{(k)}$ and make a positive definite projection if it is not positive definite, then we rescale $\Sigma^{(k)}$ so that all its diagonal elements are 1. The positive definite projection is realized via R package ``BDCoColasso". For $k \notin \mathcal{A}$, we first generate $\Omega^{(k)}$. The $(i,j)$-entry of $\Omega^{(k)}$ equals to $1.5*I(i=j)+\delta_{ij}$, where $\delta_{ij}$ is zero with probability 0.9 and is 0.2 with probability 0.1. We also symmetrize $\Omega^{(k)}$ and make a positive definite projection if it is not positive definite via R package ``BDCoColasso". Let $\Sigma^{(k)} = (\Omega^{(k)})^{-1}$ for $k\notin\mathcal{A}$. To obtain the correlation matrix, we rescale $\Sigma^{(k)}$ so that all its diagonal elements are 1.

We then sample $n$ data points $x_1,\dots,x_n$ from the Gaussian Copula distribution $\text{NPN}_p(f,\Sigma)$ and $n_k$ data points $x_1^{(k)},\dots,x_{n_k}^{(k)}$ from $\text{NPN}_p(f^{(k)},\Sigma^{(k)}),k=1,\dots,K$. For simplicity, we use the same transformation functions for the target study and the auxiliary study and for each dimension, that is, $f_1=\dots=f_p=f_1^{(k)}=\dots=f_p^{(k)}=f^0,k=1,2,\dots,K$. To sample data from the Gaussian Copula distribution, we also need $g^0 :=(f^0)^{-1}$. We consider the following two versions of $g^0$:
\begin{definition}
	(Gaussian CDF transformation). Let $g_0$ be a univariate Gaussian cumulative distribution function with mean $\mu_{g_0}$ and the standard deviation $\sigma_{g_0}:g_0(t):=\Phi(\frac{t-\mu_{g_0}}{\sigma_{g_0}})$. The Gaussian CDF transformation $g_j=(f^0)^{-1}$ for the $j$th dimension is defined as \[g_j(z_j)=\frac{g_0(z_j)-\int g_0(t)\phi((t-\mu_j)/\sigma_j)\,dt}{\sqrt{\int(g_0(y)-\int g_0(t)\phi((t-\mu_j)/\sigma_j)\,dt)^2\phi((y-\mu_j)/\sigma_j)\,dy}},\]
	where $\phi(\cdot)$ is the standard Gaussian density function and $\sigma_j=\Sigma_{jj}$.
\end{definition}
This Gaussian CDF transformation is introduced in \cite{liu2009nonparanormal}  and we set $\mu_{g_0}=0.05$ and $\sigma_{g_0}=0.4$ as in their paper.
\begin{definition}
	(Exponential transformation). The Exponential transformation $g_j=(f^0)^{-1}$ for the $j$-th dimension is defined as
	\[g_j(z_j)=\exp(z_j).\]
\end{definition}
In addition to the above two versions of transformation functions, we also consider the linear transformation (or no transformation) to investigate whether the semiparametric methods are still valid when data are truly Gaussian.

For CLIME and Copula-CLIME, we use all the data from the target study directly to estimate the correlation matrix and then estimate the precision matrix. For the other four methods, as we perform an aggregation step, we divide the target data into two folds, with sample sizes of $2n/3$ and $n/3$ respectively. We use $2n/3$ samples to compute $\widehat{\Omega}^{(\text{CL})}$ and perform the first three steps. We use the rest $n/3$ samples to perform the aggregation step. For the setting of tuning parameters, we consider $\lambda_{\text{CL}}=2c_n\sqrt{\log p/n}$, $\lambda_{\Delta}=2\sqrt{\log p/n}$ and $\lambda_{\Omega}=2c_n\sqrt{\log p/n_{\mathcal{A}}}$ where $c_n$ is selected by five fold cross validation to minimize the prediction error   defined as follows:
\begin{equation}\label{eqcv}
	\widehat{Q}(\Omega^{(a)})=\frac{1}{2p}\bigg[\text{Tr}(\hat{S}^{(t)}\Omega^{(a)}_+)-\log|\Omega^{(a)}_+|\bigg],
\end{equation}
where $\widehat{S}^{(t)}$ is the estimator of the correlation matrix $\Sigma$ using  testing set $x_i^{(t)},i=1,\dots,n_{(t)}$, $\Omega^{(a)}$ is an arbitrary graph estimator, and $\Omega^{(a)}_+$ is a symmetric positive matrix based on $\Omega^{(a)}$. Actually, (\ref{eqcv}) is defined based on the negative log-likelihood.

\begin{figure}
	\includegraphics[width=15cm]{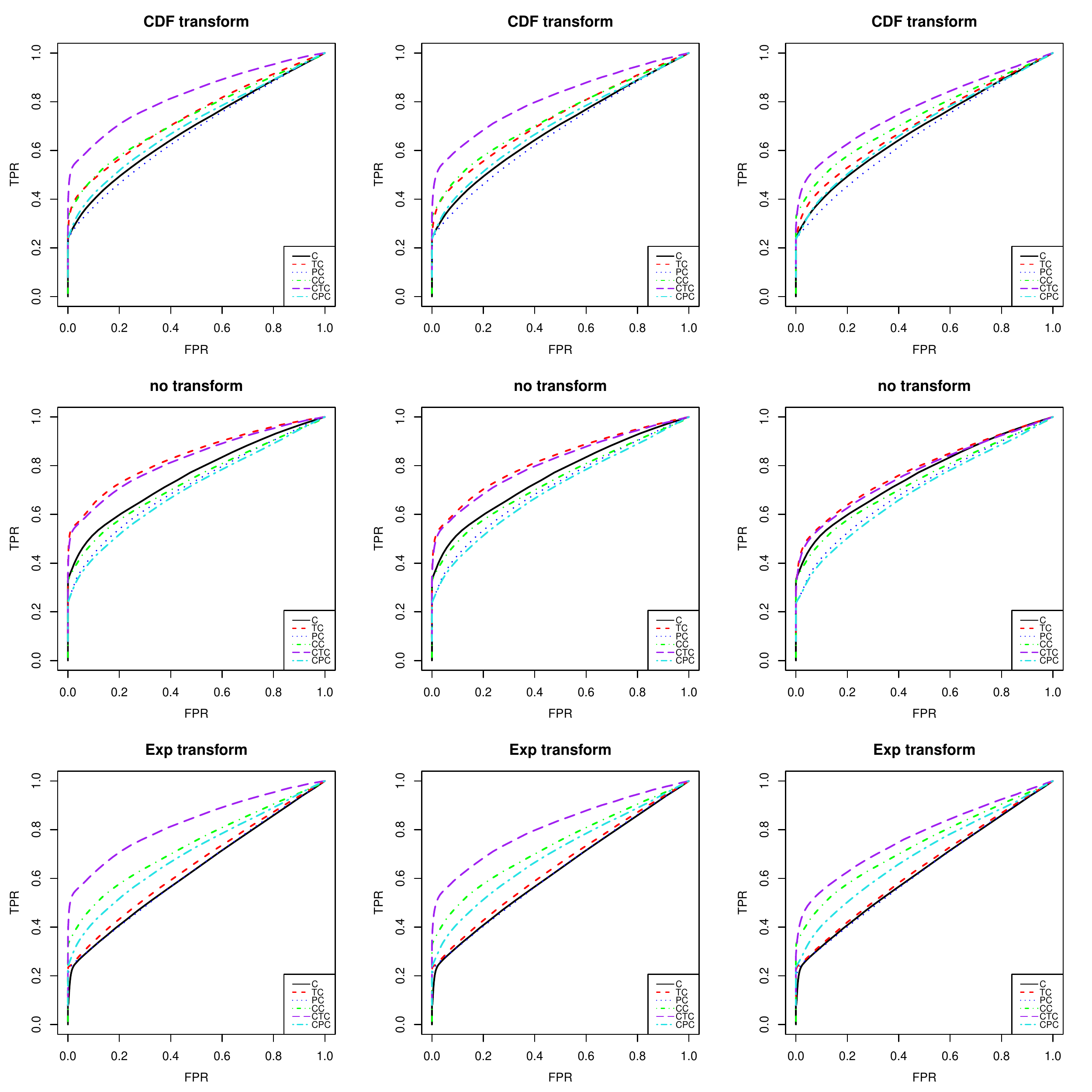}
	\fcaption{ROC curves for the Gaussian CDF, linear and exponential transformations (top, middle, bottom) using the six methods with banded graph structure, with similarity between the target study and auxiliary studies at different levels ($r=10, 20, 30$ from left to right), and $p=100,n=n_1=\dots=n_k=200,K=5,\mathcal{A}=\{1,2,3\}$.}\label{fig:1}
\end{figure}

Let $\mathcal{A}=\{1,2,3\}$, we draw averaged ROC curves over 100 trials for six methods with different transformation funtions and different types of precision matrices, as shown in Figure \ref{fig:1} and Figure \ref{fig:2}. The abscissa of ROC curve represents the false negative rate, and the ordinate represents the true positive rate.
In addition to ROC curves, we compare the estimation errors in terms of Frobenius norm as a function of the number of informative studies for six methods in different settings. The number of informative studies ranges from 1 to 5. The averaged estimation errors in Frobenius norm over 100 trials are shown in Figure \ref{fig:3} and Figure \ref{fig:4}.
\subsection{Analysis of Simulation Results}
\subsubsection{Summary of the results}
\begin{figure}
	\includegraphics[width=15cm]{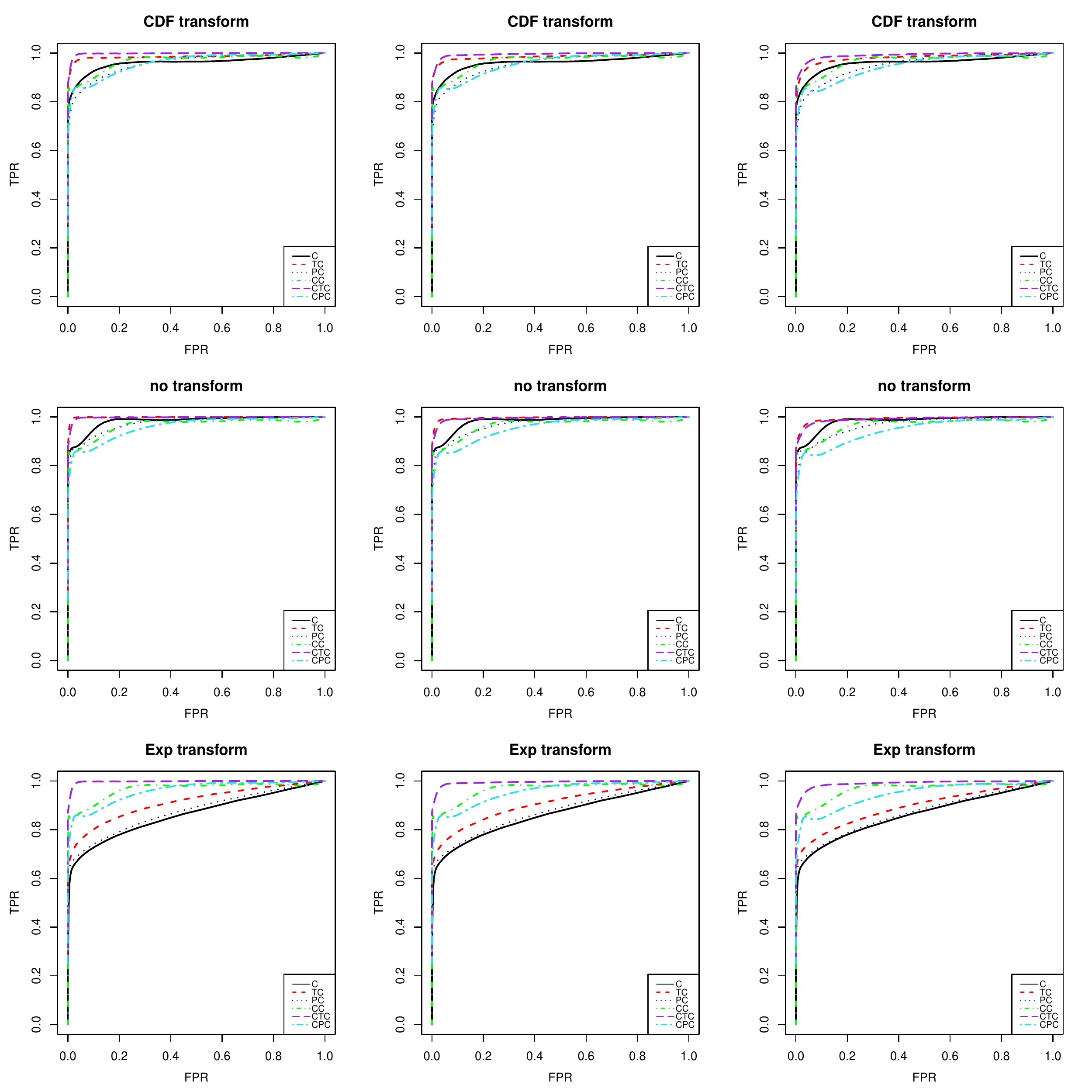}
	\fcaption{ROC curves for the Gaussian CDF, linear and exponential transformations (top, middle, bottom) using the	six methods with block diagonal graph structure, with similarity between target study and auxiliary studies at different levels ($r=10, 20, 30$ from left to right), and $p=100,n=n_1=\dots=n_k=200,K=5,\mathcal{A}=\{1,2,3\}$.}\label{fig:2}
\end{figure}

From the ROC curves in Figure \ref{fig:1} and Figure \ref{fig:2}, we see that, if we have a certain number of informative auxiliary data, Trans-CLIME estimator performs slightly better than the  Trans-Copula-CLIME estimator when the data is from Gaussian distribution. However, when the data is not from Gaussian distribution, the performance of the Trans-Copula-CLIME estimator is significantly better than the other five methods. This shows that the combination of transfer learning and semi-parametric method improves both statistical efficiency and estimation robustness. Likewise, the Copula CLIME estimator outperforms the CLIME estimator and the Copula Pooled Trans-CLIME estimator outperforms the Pooled Trans-CLIME estimator when the data are not from Gaussian, which shows the robustness of the semi-parametric methods once again.

In addition, we can also see that the performance of the  Trans-Copula-CLIME estimator is the best among the three copula-based methods, followed by Copula CLIME, and the Copula Pooled Trans-CLIME performs the worst. Similarly, for the three parametric methods (CLIME, Trans-CLIME and Pooled Trans-CLIME), the performance of the Trans-CLIME estimator is better than CLIME estimator and the performance of the Pooled Trans-CLIME estimator is the worst. These results indicate that the informative auxiliary data can improve the performance of the estimator, while the non-informative auxiliary data makes no difference to the performance of the estimator. Instead, because we only use $2n/3$ samples to compute $\widehat{\Omega}^{(\text{CL})}$, the performance of the pooled version is  even worse than CLIME or Copula CLIME estimator.

The performance of the  Trans-Copula-CLIME estimator and Trans-CLIME estimator is related to the similarity between the target study and the auxiliary studies. When the similarity level increases, the performance of the two methods also improves. Especially when $r=10$ (the most similar case in our setting), which means the difference between the target study and the auxiliary studies is negligible, these two methods are far better than the corresponding CLIME method only using samples from the single target study.

\begin{figure}
	\includegraphics[width=15cm]{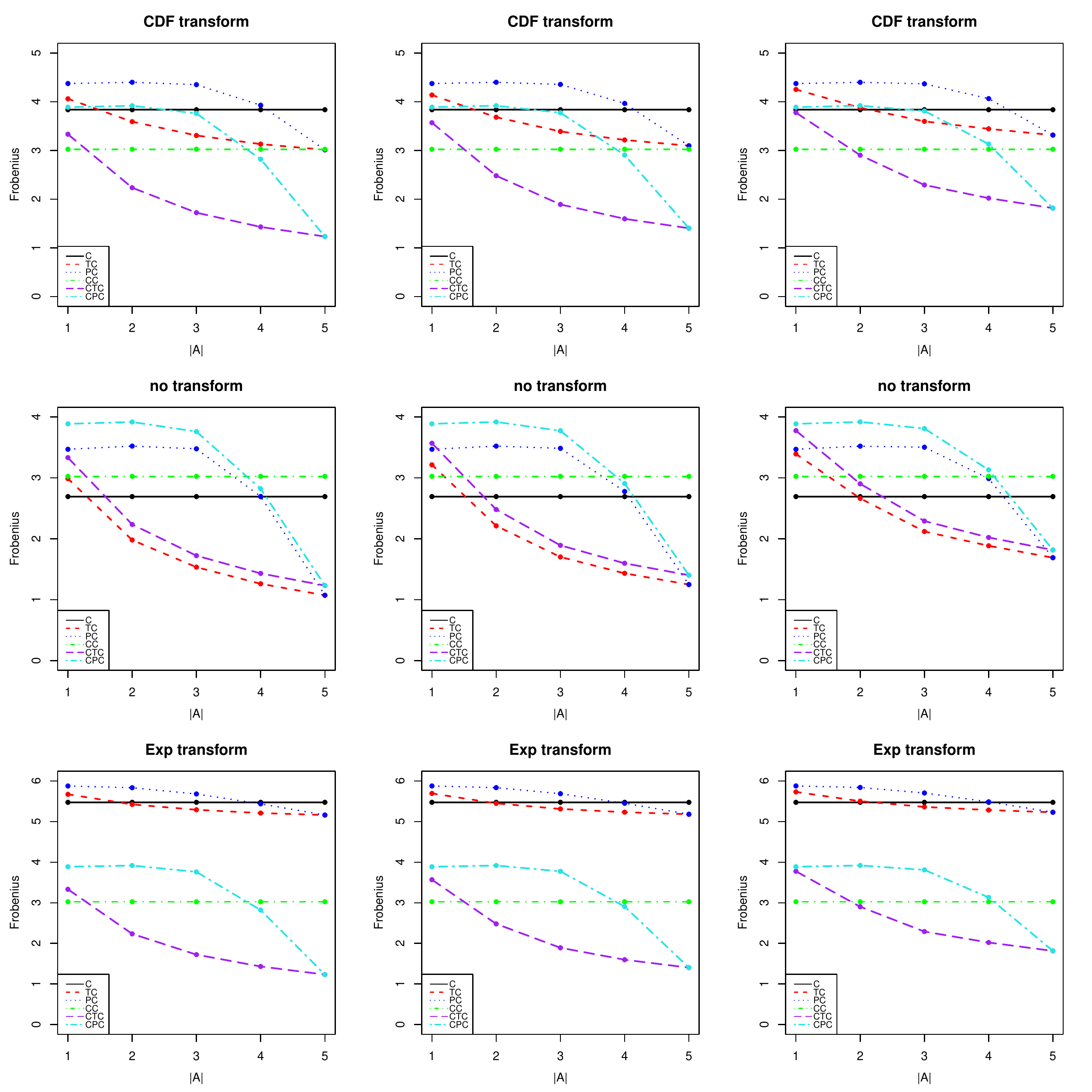}
	\fcaption{Estimation errors in Frobenius norm as a function of the number of informative studies for the Gaussian CDF, linear and exponential transformations (top, middle, bottom) using the	six methods with banded graph structure, with similarity between target study and auxiliary studies at different levels ($r=10, 20, 30$ from left to right), and $p=100,n=n_1=\dots=n_k=200,K=5$.}\label{fig:3}
\end{figure}

\begin{figure}
	\includegraphics[width=15cm]{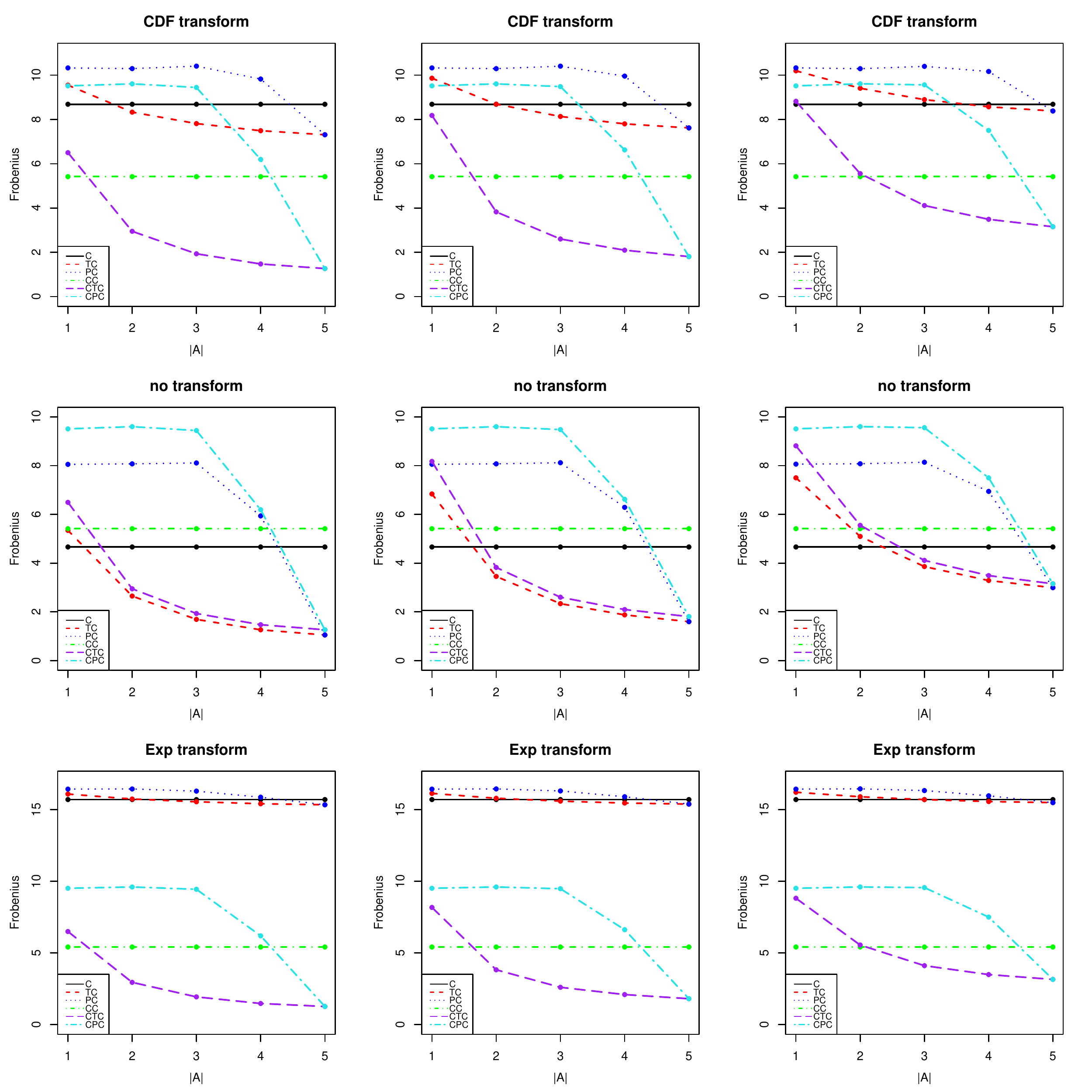}
	\fcaption{Estimation errors in Frobenius norm as a function of the number of informative studies for the Gaussian CDF, linear and exponential transformations (top, middle, bottom) using the six methods with block diagonal graph structure, with similarity between target study and auxiliary studies at different levels ($r=10, 20, 30$ from left to right), and $p=100,n=n_1=\dots=n_k=200,K=5$.}\label{fig:4}
\end{figure}

From the estimation errors curves in terms of Frobenius norm shown in Figure \ref{fig:3} and Figure \ref{fig:4}, we can draw similar conclusions. Besides, as the amount of informative auxiliary data increases, the estimation errors of the  Trans-Copula-CLIME estimator and the Trans-CLIME estimator decrease, which validates our theoretical results. When the number of informative auxiliary studies is small, estimators based on pooled version do not work well. This is because $K-|\mathcal{A}|$ non-informative studies are used in the pooled version, which deteriorates the empirical performance.
In the following, we make further analysis according to ROC curves in Figure \ref{fig:1} and Figure \ref{fig:2}.

\subsubsection{Non-Gaussian data}
To compare the graph estimation performance of two procedures $A$ and $B$, in the following we use $A>B$ to represent that $A$ is better than $B$; $A \gg B$ means that $A$ is significantly better than $B$; $A \approx B$ means that $A$ and $B$ have similar performance.

The results of the comparison of the six methods with different precision matrices and transformation functions are as follows:\\\vspace{0.5em}
$\bullet$ Non-Gaussian data with banded matrix and CDF transformation: CTC$\gg$CC$>$TC$>$CPC$>$C$>$PC.\\ \vspace{0.5em}
$\bullet$ Non-Gaussian data with banded matrix and Exp transformation: CTC$\gg$CC$>$CPC$>$TC$>$C$\approx$PC.\\\vspace{0.5em}
$\bullet$ Non-Gaussian data with block diagonal matrix and CDF transformation: CTC$>$TC$\gg$CC$\approx$C$>$CPC$\approx$PC.\\\vspace{0.5em}
$\bullet$ Non-Gaussian data with block diagonal matrix and Exp transformation: CTC$\gg$CC$>$CPC$>$TC$>$PC$>$C.\vspace{0.5em}

From the Gaussian CDF transformation and exponential transformation ROC curves in Figure \ref{fig:1} and Figure \ref{fig:2}, we see that, the three copula-based methods are better than the corresponding parametric methods on the whole. The performance of the  Trans-Copula-CLIME estimator is the best among the six methods and significantly better than Trans-CLIME estimator. Even the performance of the Copula CLIME estimator is better than Trans-CLIME estimator in some settings. Especially when the transformation function is exponential, the Copula Pooled Trans-CLIME estimator also outperforms Trans-CLIME estimator. This shows that the distribution of the data is very important for the choice of different methods. When the data comes from a non-Gaussian distribution, it is better to use semi-parametric approach with the single target study than to use parametric method with informative auxiliary data.

\subsubsection{Gaussian data}
The results of the comparison of the six methods with different precision matrices are as follows:\\\vspace{0.5em}
$\bullet$ Gaussian data with banded matrix: TC$>$CTC$\gg$C$>$CC$>$PC$>$CPC.\\\vspace{0.5em}
$\bullet$ Gaussian data  block diagonal matrix: CTC$\approx$TC$\gg$C$>$CC$\approx$PC$>$CPC.\vspace{0.5em}

From linear transformation curves in Figure \ref{fig:1} and Figure \ref{fig:2}, we see that, contrary to the results for non-Gaussian data, the performance of the Trans-CLIME estimator is slightly better than the Trans-Copula-CLIME estimator. The performance of the CLIME estimator is better than Copula CLIME estimator. The Pooled Trans-CLIME estimator outperforms Copula Pooled Trans-CLIME estimator. However, the ROC curves of Trans-CLIME estimator and  Trans-Copula-CLIME estimator are pretty close. Especially when the precision matrix has block diagonal structure, Trans-CLIME estimator and  Trans-Copula-CLIME estimator have similar performance, which shows the former does not show much advantage.

In summary, the simulation results show when the data does not satisfy the Gaussian assumption, the  Trans-Copula-CLIME estimator still retains advantageous and is the best one among the six methods. This suggests that Trans-Copula-CLIME has higher statistical efficiency and shows estimation robustness in a wider range of applications. The higher the similarity level between the auxiliary studies and the target study, the more auxiliary samples, the better the estimator with transfer learning performs. However, the pooled versions are the worst among the six methods, indicating that when the auxiliary studies differs greatly from the target study,  transfer learning does not work well. In practice, if we have informative auxiliary data, we can choose the more robust  Trans-copula-CLIME to estimate the graph.

\section{Real fMRI data related to ADHD}\label{sec:realexa}
In this section, we apply the proposed algorithm to study the brain connectivity in patients with attention deficit hyperactivity disorder (ADHD) using the dataset from ADHD-200 Global Competition. The dataset includes demographical information and resting state functional magnetic resonance imaging (fMRI) of nearly one thousand children and adolescents, including both combined types of ADHD and typically developing controls (TDC). We focus our analysis on the fMRI data from Beijing and New York sites, with Beijing site dataset selected as the target study and the New York site data viewed as the auxiliary study. We preprocessed the dataset with steps including correction, smoothing, denoising, quality control and deletion of missing values, finally resulting in a dataset consisted of 74 combined ADHD subjects and 109 TDC subjects in Beijing site and 96 combined ADHD subjects and 91 TDC subjects in New York site. Each brain image was parcellated into 116 regions of interest (ROI) using the Anatomical Automatic Labeling (AAL) atlas. The time series of voxels within the same ROI were then averaged.

\begin{figure}
	\includegraphics[width=15cm]{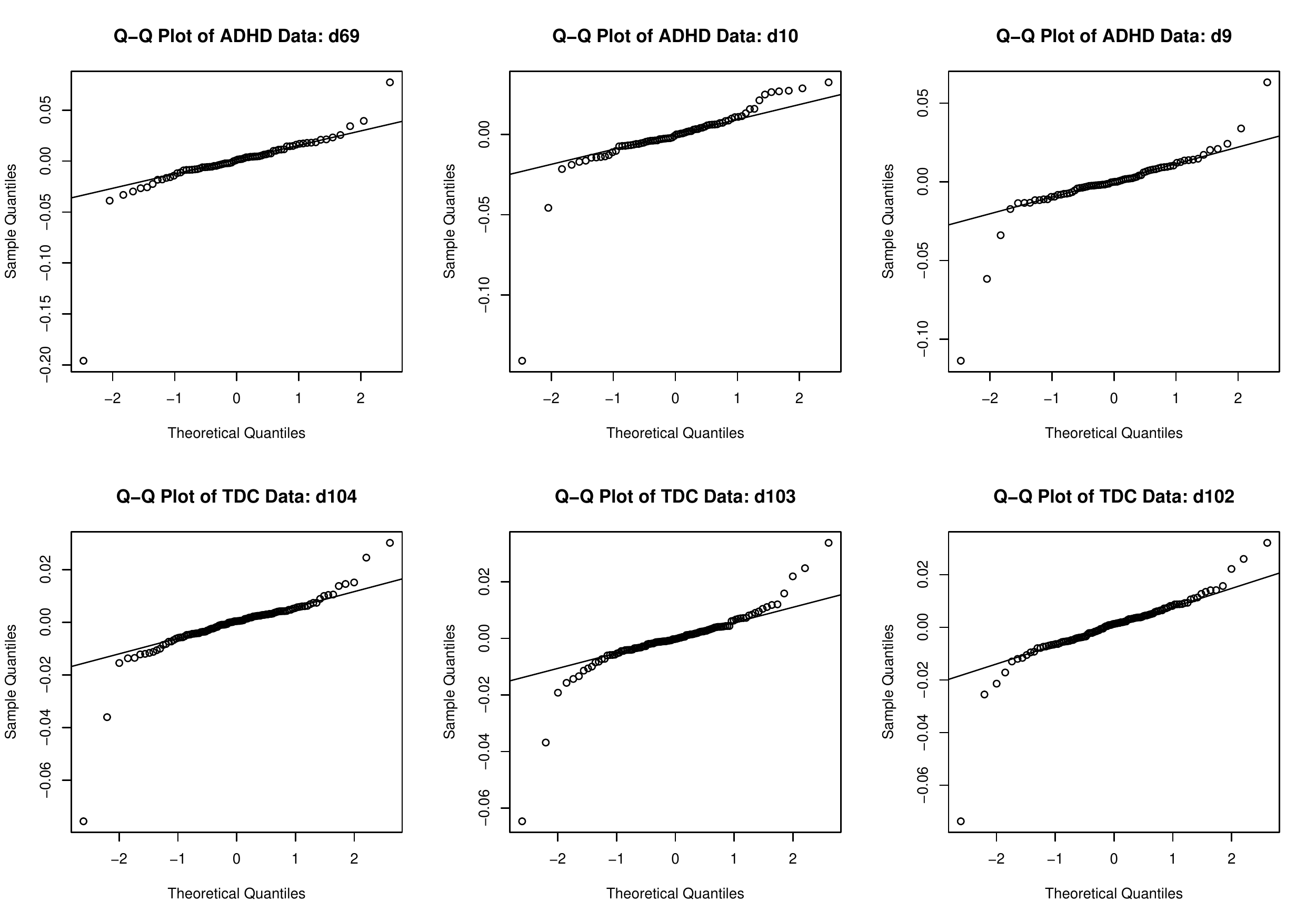}
	\fcaption{Q-Q plots of ADHD subjects in Beijing (the first line) and TDC subjects in Beijing (the second line). "d" stands for dimension, for example, "d69" stands for 69th dimension.}\label{fig:5}
\end{figure}

First, we use Henze-Zirkler's multivariate normality test and Royston's multivariate normality test in R package ``MVN" (\cite{Selcuk2014MVN}) to check the normality of data and reject the null hypothesis (both $p$-values is zero or almost zero). Then, we use shapiro-Wilk test to check the normality of each dimension, and select some dimensions to draw the Q-Q plots, as shown in Figure \ref{fig:5}. The top panel shows the Q-Q plots of ADHD subjects in Beijing site, and the bottom panel shows the Q-Q plots of TDC subjects in Beijing site. As can be seen from Figure \ref{fig:5}, our data seriously violates the normality assumption. Hence, it is reasonable to use  Trans-Copula-CLIME to analyze the data in the following.

\begin{figure}[H]
	\centering
	\subfigure[]{
		\begin{minipage}[t]{0.3\linewidth}
			\centering
			\includegraphics[width=1\textwidth]{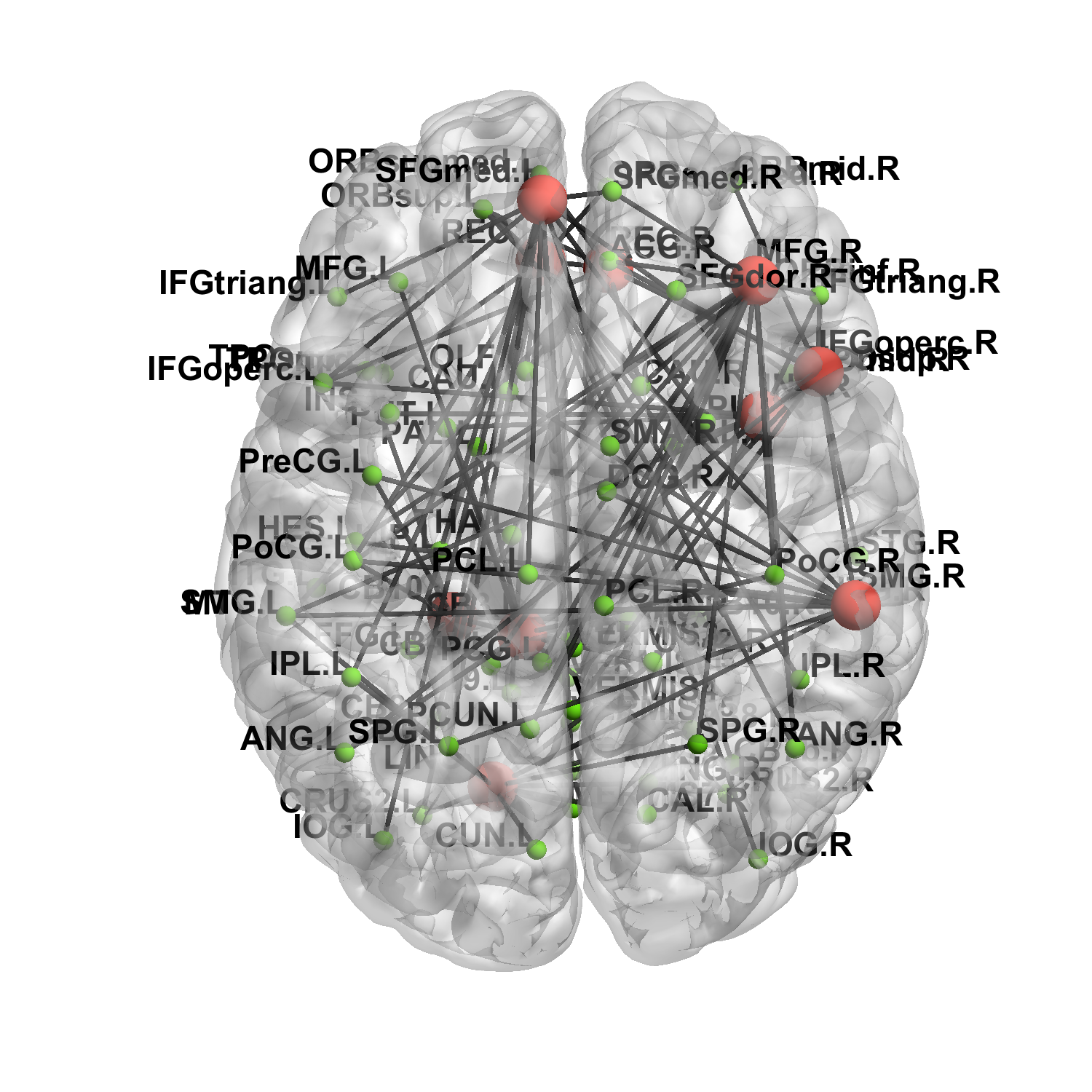}
		\end{minipage}
	}
	\subfigure[]{
		\begin{minipage}[t]{0.3\linewidth}
			\centering
			\includegraphics[width=1\textwidth]{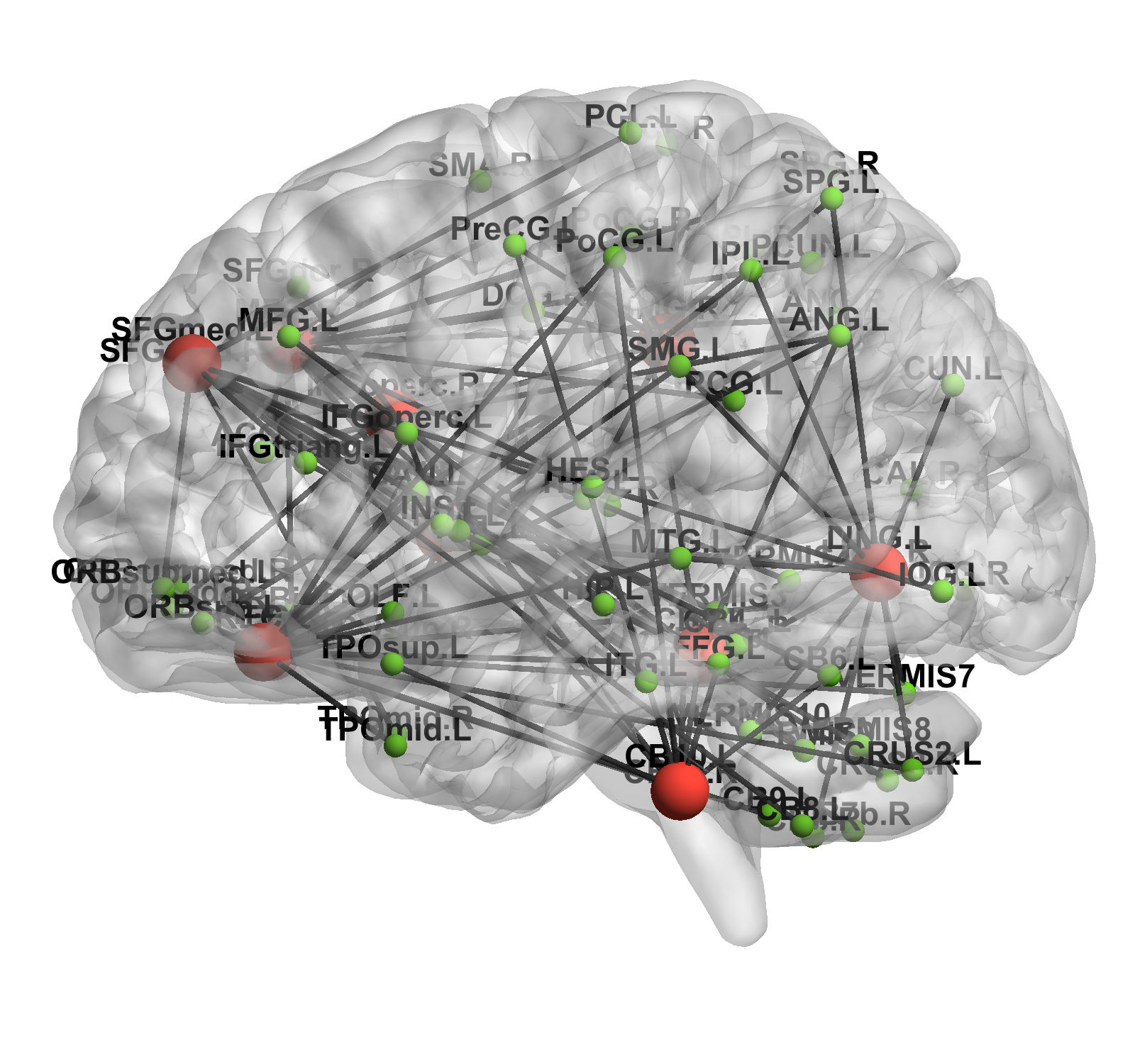}
		\end{minipage}
	}
	\subfigure[]{
		\begin{minipage}[t]{0.3\linewidth}
			\centering
			\includegraphics[width=1\textwidth]{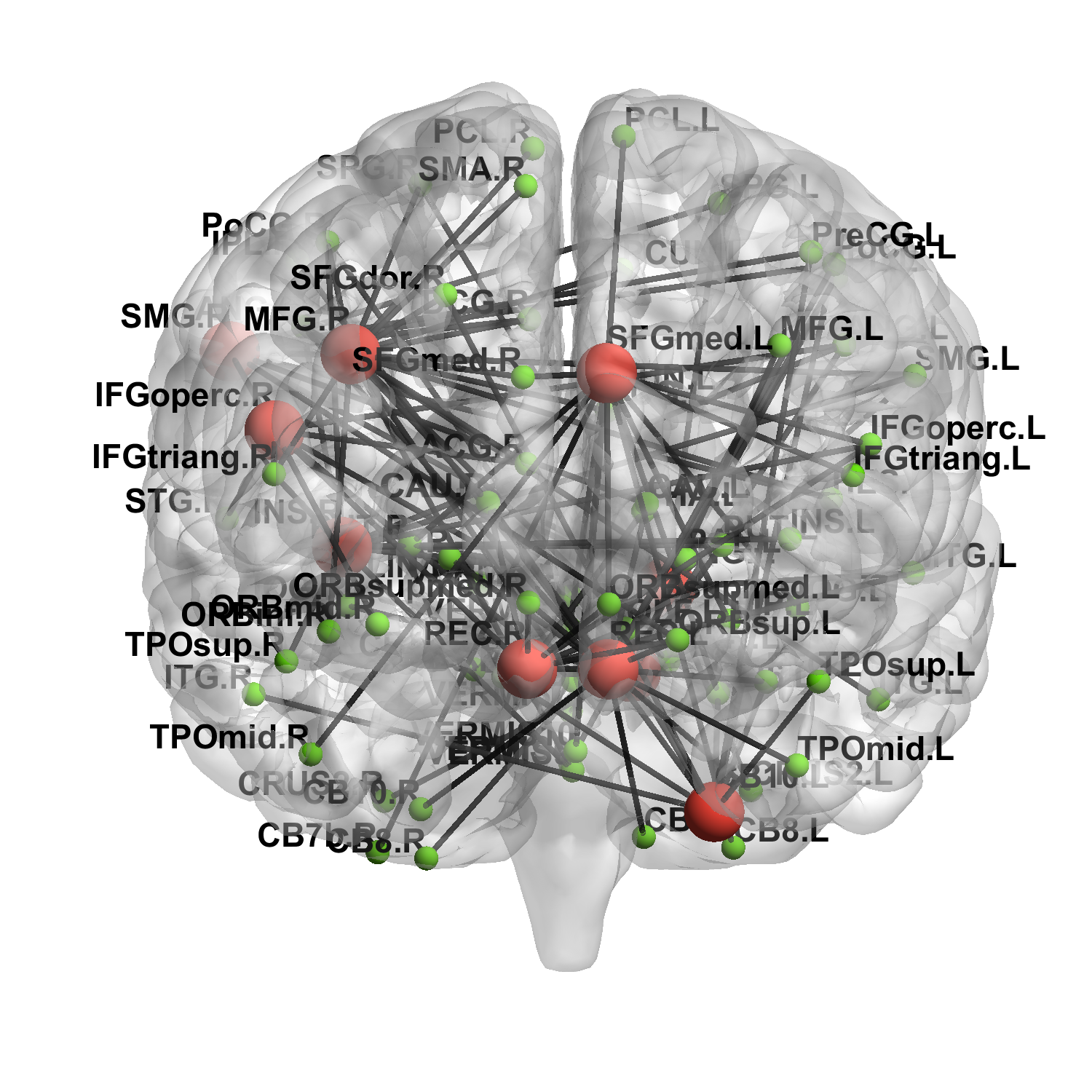}
		\end{minipage}
	}
	
	\subfigure[]{
		\begin{minipage}[t]{0.3\linewidth}
			\centering
			\includegraphics[width=1\textwidth]{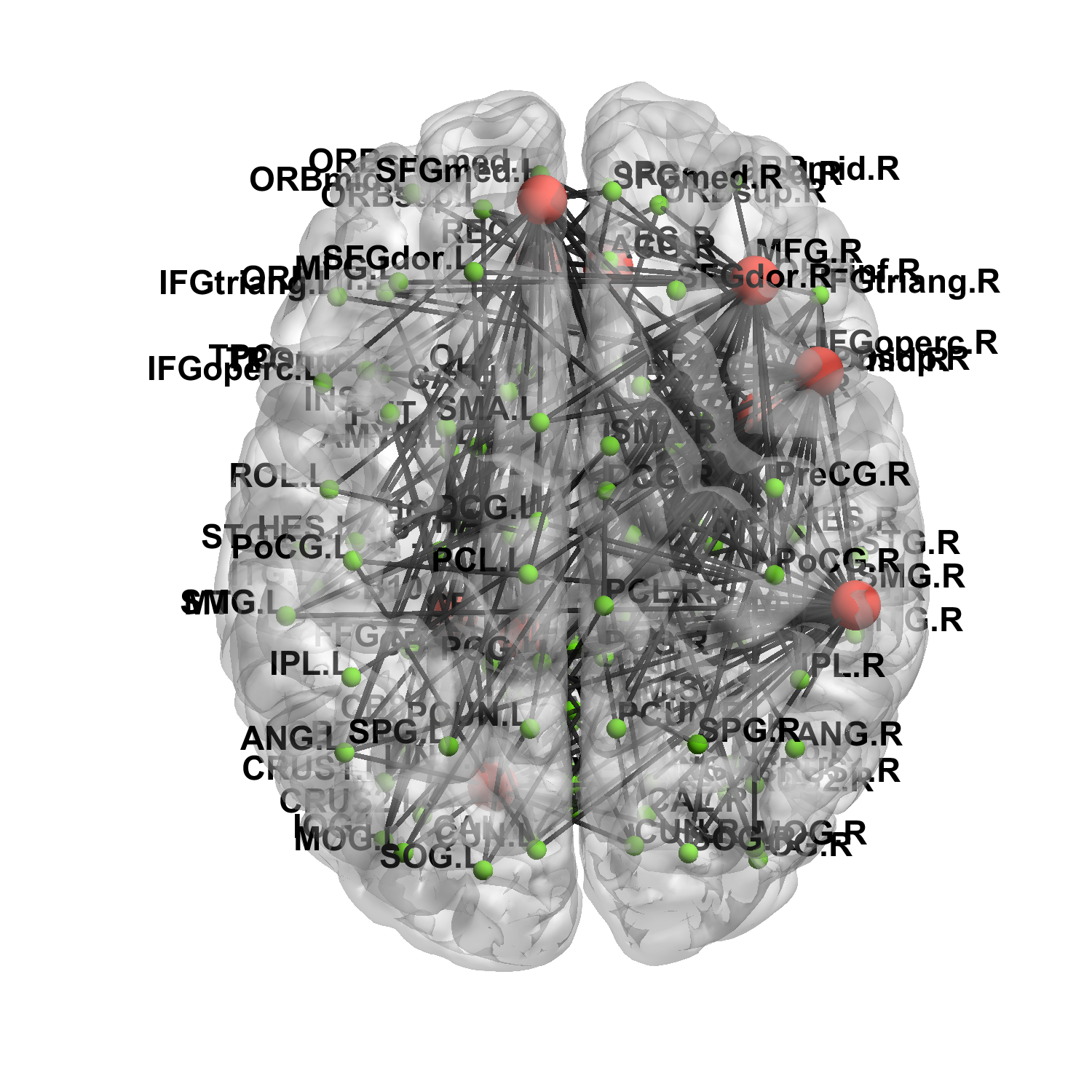}
		\end{minipage}
	}
	\subfigure[]{
		\begin{minipage}[t]{0.3\linewidth}
			\centering
			\includegraphics[width=1\textwidth]{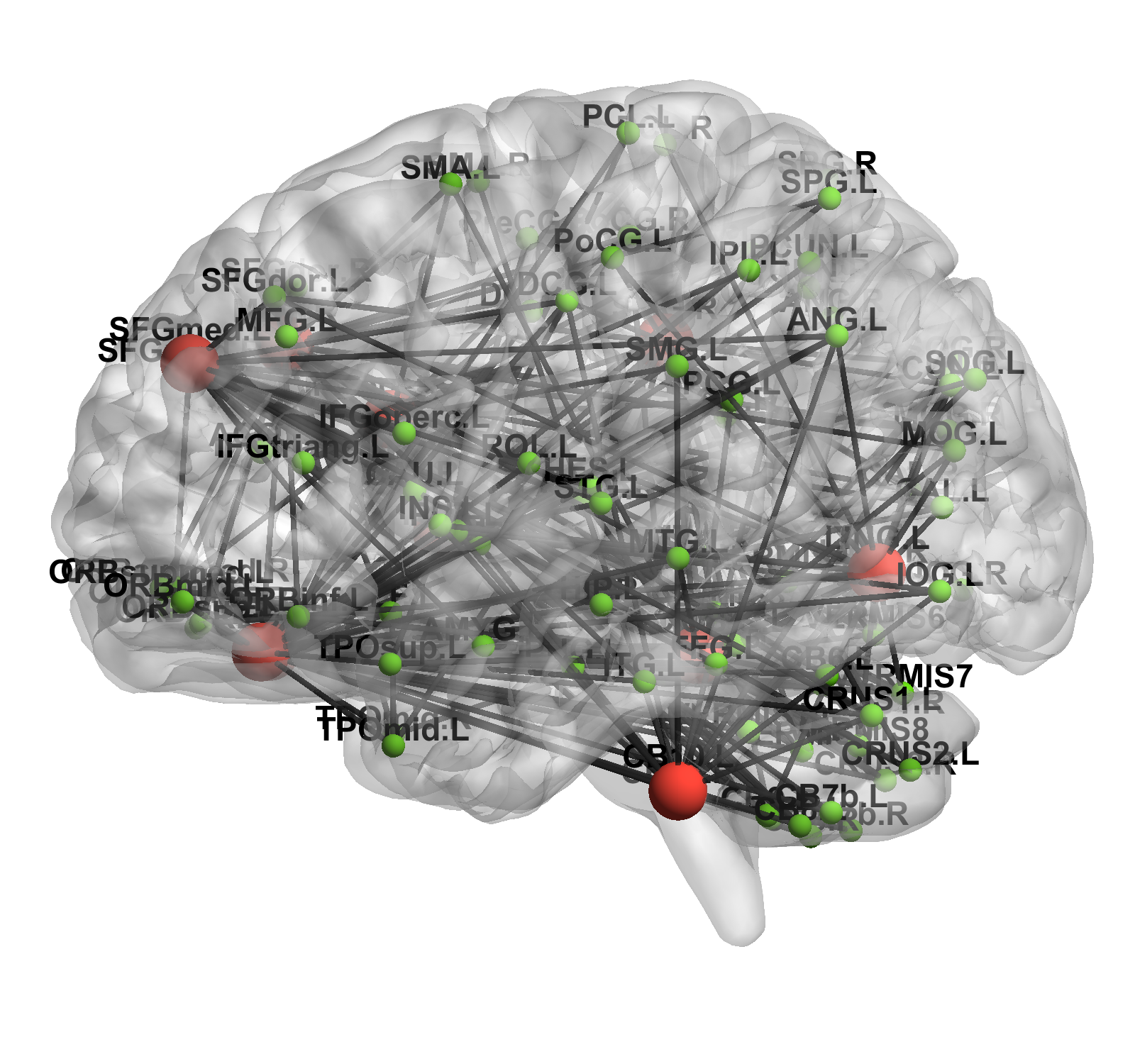}
		\end{minipage}
	}
	\subfigure[]{
		\begin{minipage}[t]{0.3\linewidth}
			\centering
			\includegraphics[width=1\textwidth]{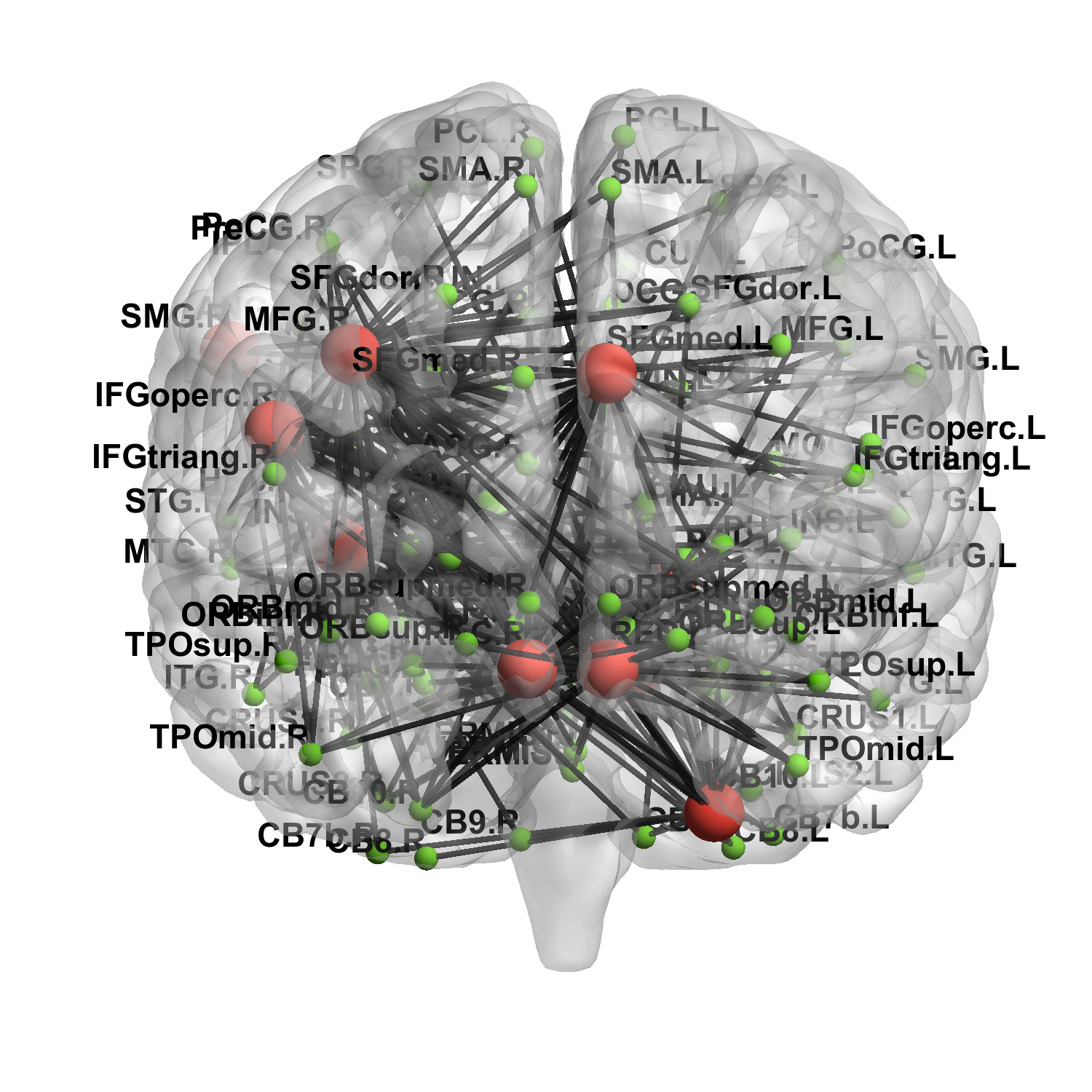}
		\end{minipage}
	}
	
	\subfigure[]{
		\begin{minipage}[t]{0.3\linewidth}
			\centering
			\includegraphics[width=1\textwidth]{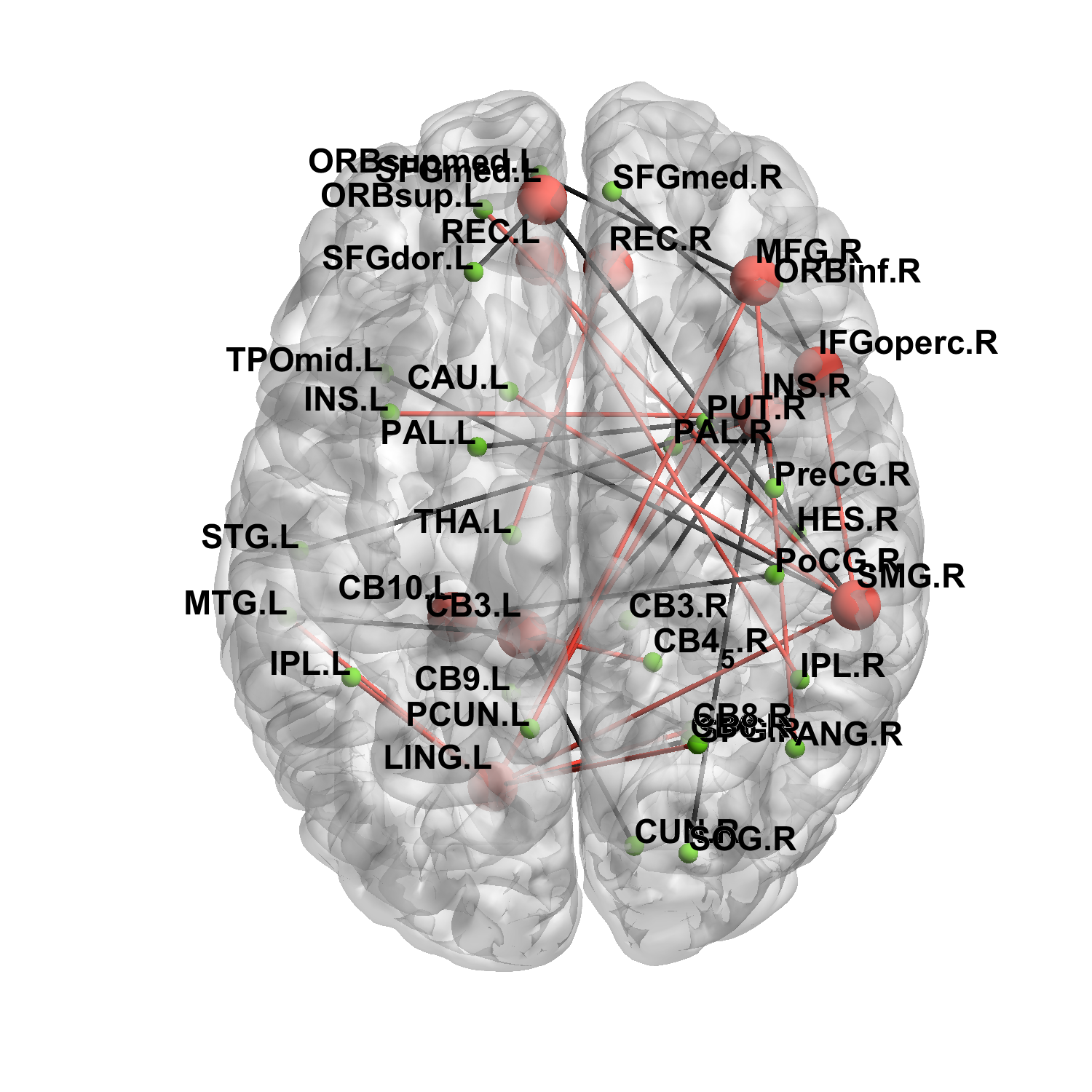}
		\end{minipage}
	}
	\subfigure[]{
		\begin{minipage}[t]{0.3\linewidth}
			\centering
			\includegraphics[width=1\textwidth]{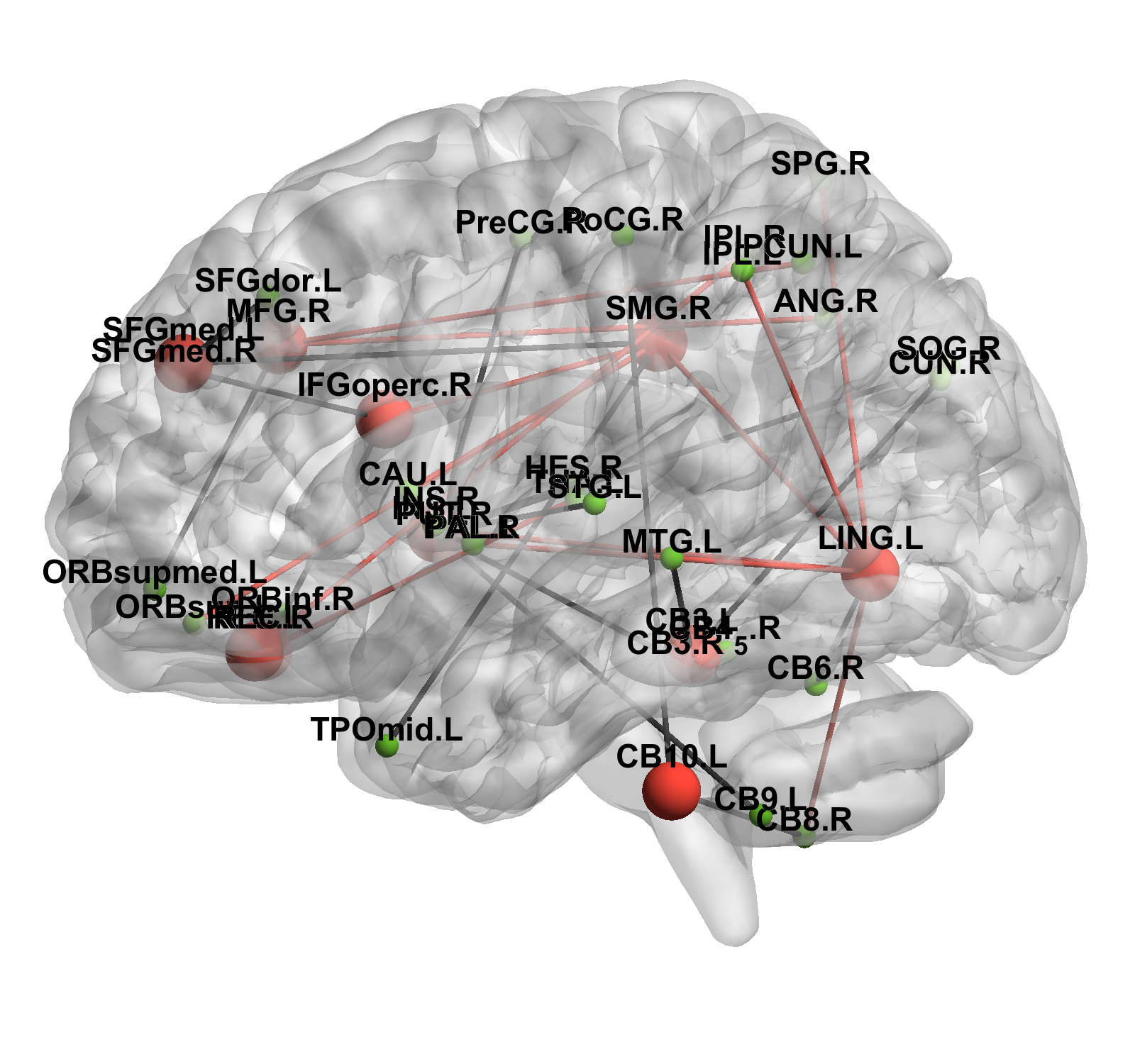}
		\end{minipage}
	}
	\subfigure[]{
		\begin{minipage}[t]{0.3\linewidth}
			\centering
			\includegraphics[width=1\textwidth]{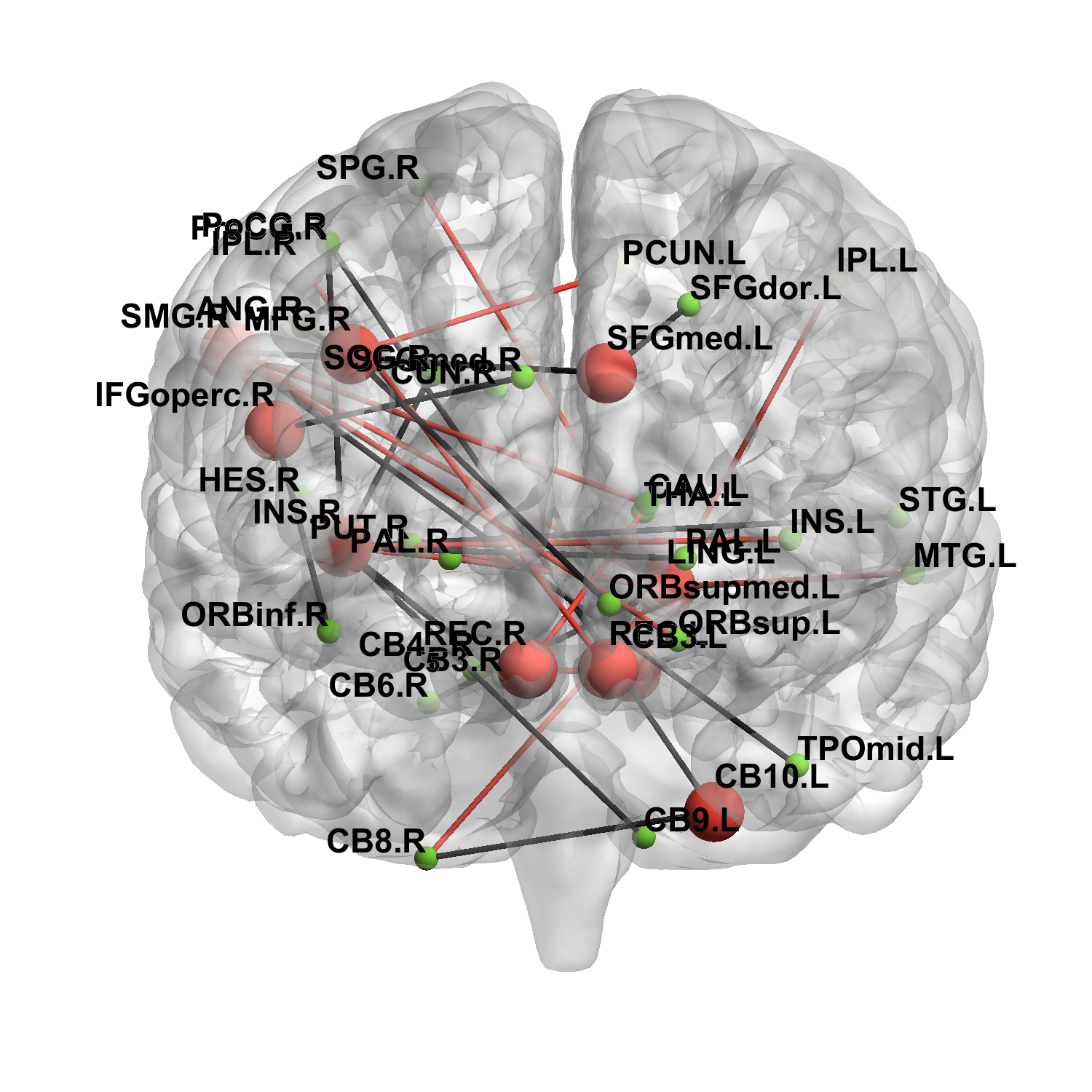}
		\end{minipage}
	}
	\centering
	\caption{Graphical models estimated by  Trans-Copula-CLIME for the ADHD resting-state fMRI data. (a)-(c) show the connections between the ten brain regions and other brain regions of ADHD subjects, (d)-(f) show the connections between these ten brain regions and other brain regions of TDC subjects, and (g)-(i) show the top 10\% differential edges between the two graphical models. The red dots in  (a)-(i) represent the ten brain regions with the greatest differences between the two graphical models, while the green dots represent the other brain regions. The red edges in  (g)-(i) exist in the graph of ADHD subjects, but does not exist in the graph of TDC subjects; the black edges exist in the graph of TDC subjects, but does not exist in the graph of ADHD subjects.}\label{fig:6}
\end{figure}

We apply  Trans-Copula-CLIME to estimate the brain connectivity of ADHD subjects from Beijing and use ADHD subjects from New York as the auxiliary study. Similarly, we apply  Trans-Copula-CLIME to estimate the brain connectivity of TDC subjects from Beijing and use TDC subjects from New York as the auxiliary study. Then we identify ten brain regions with the greatest differences between the two estimated graphical models and draw Figure \ref{fig:6} using the BrainNet Viewer software (\cite{xia2013brainnet}). In Figure \ref{fig:6}, (a)-(c) show the connections between the ten brain regions and other brain regions in the graphical model of ADHD subjects from the axial view, coronal view, and sagittal view, respectively; (d)-(f) show the connections between these ten brain regions and other brain regions in the graphical model of TDC subjects from the axial view, coronal view, and sagittal view, respectively, (g)-(i) show the top 10\% differential edges between the two graphical models between these ten brain regions and other brain regions from the axial view, coronal view, and sagittal view, respectively.

As shown in Figure \ref{fig:6}, the brain regions related to ADHD are mainly located in middle frontal gyrus, gyrus rectus, insula, supramarginal gyrus, inferior frontal gyrus opercular part, lingual gyrus, cerebellum and superior frontal gyrus medial. Many developmental disorders, such as ADHD and autism, are associated with cerebellar dysfunction, which may have long-term effects on motor, cognitive and emotional regulation \citep{stoodley2016cerebellum}. \cite{ivanov2014cerebellar} revealed that compared with the control group, adolescents with ADHD had smaller volumes in subregions of the cerebellum. It is reported that the middle frontal gyrus is involved in the storage and processing of working memory \citep{leung2002sustained}, and the lateral prefrontal cortex composed of the middle frontal gyrus is involved in dynamic cognitive control processes \citep{ridderinkhof2004neurocognitive}. \cite{ding2021identification} found that the functional connectivity between the cerebellum and middle frontal gyrus was enhanced in ADHD children, which may lead to the disorder of the connection network, resulting in attention deficit in ADHD children. It was found that the right inferior frontal gyrus plays an important role in response inhibition and attentional control \citep{hampshire2010role}. \cite{cubillo2010reduced} revealed that patients with ADHD show fewer functional connectivity between the left and right inferior frontal gyrus, caudate/ thalamus, cingulate gyrus, and temporal/parietal regions during a response inhibition task than that in controls. \cite{depue2010behavioral} found that there are morphological differences in inferior prefrontal regions between young people with ADHD and healthy controls. Patients with ADHD had smaller gray matter volume in the right inferior frontal gyrus, and this was associated with poorer behavioral performance. The insula plays a complex and critical role in human cognition, involving attention, emotional awareness, decision-making and cognitive control \citep{craig2009emotional}. \cite{lopez2012reduced} showed that the anterior insula gray matter volume of adolescents with ADHD was smaller than that of the healthy group. They further found that the left anterior insula gray matter was associated with oppositional symptoms, while the right anterior insula gray matter was associated with attention problems and inhibition. The lingual gyrus is involved in visual speech processing and word processing \citep{bogousslavsky1987lingual,mechelli2000differential}. \cite{ko2018altered} found that adults with ADHD responded to the dual-task effect with higher levels of activation in the left tongue region. \cite{dibbets2010differential} found that ADHD patients have different activation patterns in brain frontal striatum circuit during the performance of an executive control task compared to the control group. Adults with ADHD showed greater activity in several brain regions such as the middle temporal gyrus, lingual gyrus and insula during task switching compared to healthy controls.
\section{Conclusion}
We proposed  Trans-Copula-CLIME estimator to estimate an undirected graphical model by transfer learning information from auxiliary studies without Gaussian assumption. We obtain the asymptotic properties of the estimator in high-dimensional setting, which show that as long as we have sufficient samples from related auxiliary studies similar to target study, the transfer learning brings great advantage. Simulation studies validate the theoretical results and also show
that Trans-Copula-CLIME estimator has better performance especially when data are not from Gaussian
distribution.  However, the theoretical properties of the estimator after aggregation step is still not clear, which is listed as our future research direction.

\clearpage

\begin{center}
	APPENDIX: PROOFS OF MAIN THEOREM
\end{center}

\begin{appendices}
	
	\section{Proofs of Main Theorems}
	
	To prove Theorem \ref{t1}, we first present some useful lemmas. Lemma \ref{CvgS} is adapted from Theorem 4.2 in \cite{liu2012high}. Lemma \ref{CvgSA} establishes the convergence rate of $\widehat{S}^{\mathcal{A}}$ in element-wise matrix  infinity norm. Lemma \ref{REC} and Lemma \ref{Cvgrj} provide two critical results in the proof of Theorem \ref{t1}. Lemma \ref{CgvCL} establishes the convergence rate of each column of the CLIME estimator in terms of $\ell_1$ and $\ell_2$ norms.
	\subsection{Some useful lemmas}
	\begin{lemma}\label{CvgS}(Convergence rate of $\widehat{S}$). For any $n>1$, we have
		$$\mathbb{P}\bigg(\|\widehat{S}-\Sigma\|_{\max} \geqslant c_{1} \sqrt{\frac{\log p}{n}} \bigg)\leqslant \exp (-c_{2} \log p)$$
	\end{lemma}
	\begin{lemma}\label{CvgSA}(Convergence rate of $\widehat{S}^{\mathcal{A}}$)
		$$\mathbb{P}\bigg(\|\widehat{S}^{\mathcal{A}}-\Sigma^{\mathcal{A}}\|_{\max}\geqslant c_1\sqrt{\frac{\log p}{n_\mathcal{A}}}\bigg) \leqslant \exp (-c_2 \log p)$$
	\end{lemma}
	\begin{lemma}\label{REC}
		For $\Sigma$ and its estimate $\widehat{S}$, if $\|\Sigma-\widehat{S}\|_{\infty,\infty}\leqslant\tilde{\lambda}$, then $\forall u\in \mathcal{C}(S,\alpha)$, we have
		$$u^{\top}\widehat{S}u \geqslant u^{\top}\Sigma u -\tilde{\lambda}s(1+\alpha)^2\|u\|_2^2$$ where $\mathcal{C}(S,\alpha):=\{u\in\mathbb{R}^p|\ \|u_{S^c}\|_1\leqslant\alpha\|u_S\|_1\}$ and here $S$ is a support set with cardinality at most $s$.
	\end{lemma}
	\begin{lemma}\label{Cvgrj}
		Under the conditions of Theorem 2.1, when $h \lesssim s \sqrt{\log p / n}$ and $n_{\mathcal{A}} \gtrsim n$, we have $$
		\mathbb{P}\bigg(\max _{j}\|r_{j}(\widehat{\Delta}^{\mathcal{A}})-r_{j}(\Delta^{\mathcal{A}})\|_{2}^{2} \geqslant c_{1} h \delta_{n}\bigg) \leqslant \exp (-c_{2} \log p)+\exp (-c_{3} n)
		$$
		for some positive constants $c_{1}, c_{2}$ and $c_{3}$.
	\end{lemma}
	\begin{lemma}\label{CgvCL}
		(Convergence rate of CLIME). Under the conditions of Theorem \ref{t1}, we have
		$$
		\begin{aligned}
			&\mathbb{P}\bigg(\|\widehat{\Omega}_j^{(\text{CL})}-\Omega_j\|_{1} \geqslant c_{1} s \sqrt{\frac{\log p}{n}}\bigg) \leqslant \exp (-c_{2} \log p) \\
			&\mathbb{P}\bigg(\|\widehat{\Omega}_j^{(\text{CL})}-\Omega_j\|_{2}^{2} \geqslant c_{1} s \frac{\log p}{n}\bigg) \leqslant \exp (-c_{2} \log p)
		\end{aligned}
		$$
		for some positive constants $c_{1}, c_{2}$ and $c_{3}$.
	\end{lemma}
	
	\subsection{Proof of Theorem \ref{t1}}	
	\begin{proof} The main idea of the proof are similar to lemma A.2 in \cite{li2020transfer}, but there are some changes to the notation and proof details. We first establish the convergence rate of $\widehat{\Omega}^L$, where $\widehat{\Omega}^L$ is an estimate of $\Omega$ using method based on Lasso. Then we establish the convergence rate of $\widehat{\Omega}$ based on $\widehat{\Omega}^{L}$.
		
		By Lemma 1 in \cite{cai2011constrained}, problem (\ref{opS2}) can be further casted as $p$ separate minimization problems and for $1 \leqslant j \leqslant p$,
		\begin{equation}\label{opS2p}
			\begin{aligned}
				&\widehat{\Omega}_{j}=\underset{\omega}{\arg \min }\|\omega\|_{1} \\
				&\quad \text { subject to }\|\widehat{S}^{\mathcal{A}} \omega-(e_{j}+r_{j}(\widehat{\Delta}^{\mathcal{A}}))\|_{\infty} \leqslant \lambda_{\Omega}.
			\end{aligned}
		\end{equation}
		We consider the Lasso version for the $j$-th column
		\begin{equation}\label{oplasso}
			\widehat{\Omega}_{j}^{L}=\underset{\omega}{\arg \min } \frac{1}{2} \omega^{\top} \widehat{S}^{A} \omega-\omega^{\top}(e_{j}+r_{j}(\widehat{\Delta}^{\mathcal{A}}))+\lambda_{\Omega}\|\omega\|_{1}.
		\end{equation}
		According to the above formula, we have oracle inequality:
		\begin{equation}\label{oi}
			\begin{aligned}
				\frac{1}{2}&(\widehat{\Omega}_{j}^{L})^{\top}\widehat{S}^{A} \widehat{\Omega}_{j}^{L}-(\widehat{\Omega}_{j}^{L})^{\top}(e_{j}+r_{j}(\widehat{\Delta}^{\mathcal{A}}))+\lambda_{\Omega}\|\widehat{\Omega}_{j}^{L}\|_{1}\\
				&\leqslant \frac{1}{2}(\Omega_{j})^{\top}\widehat{S}^{\mathcal{A}} \Omega_{j}-\Omega_{j}^{\top}(e_{j}+r_{j}(\widehat{\Delta}^{\mathcal{A}}))+\lambda_{\Omega}\|\Omega_{j}\|_{1}.
			\end{aligned}
		\end{equation}
		Let $\Omega^{*}=\Omega+\Omega^{\mathcal{A}}(\widehat{\Delta}^{\mathcal{A}}-\Delta^{\mathcal{A}})^{\top}$. We view $\Omega^{*}_j$ as the true parameter and view $\Omega_{j}$ as a spare approximation of $\Omega^{*}_j$ in the following proof.
		With $\Omega^{*}_j$, the oracle inequality (\ref{oi}) can be transformed into the following form step by step:
		$$\begin{aligned}
			&\frac{1}{2}(\widehat{\Omega}_{j}^{L})^{\top}\widehat{S}^{\mathcal{A}} \widehat{\Omega}_{j}^{L}+\frac{1}{2}(\widehat{\Omega}_{j}^{*})^{\top}\widehat{S}^{\mathcal{A}} \widehat{\Omega}_{j}^{*}-(\Omega_j^*)^{\top}\widehat{S}^{\mathcal{A}} \widehat{\Omega}_{j}^{L}-(\widehat{\Omega}_{j}^{L})^{\top}(e_{j}+r_{j}(\widehat{\Delta}^{\mathcal{A}}))+\lambda_{\Omega}\|\widehat{\Omega}_{j}^{L}\|_{1}\\
			&\text{\quad}\leqslant \frac{1}{2}\Omega_{j}^{\top}\widehat{S}^{\mathcal{A}} \Omega_{j}+\frac{1}{2}(\widehat{\Omega}_{j}^{*})^{\top}\widehat{S}^{\mathcal{A}}\widehat{\Omega}_{j}^{*}-(\Omega_j^*)^{\top}\widehat{S}^{\mathcal{A}} \Omega_j+(\Omega_j^*)^{\top}\widehat{S}^{\mathcal{A}} \Omega_j\\
			&\text{\qquad}-(\Omega_j^*)^{\top}\widehat{S}^{\mathcal{A}} \widehat{\Omega}_{j}^{L}-\Omega_{j}^{\top}(e_{j}+r_{j}(\widehat{\Delta}^{\mathcal{A}}))+\lambda_{\Omega}\|\Omega_{j}\|_{1}\\
			\Longrightarrow& \frac{1}{2} (\widehat{\Omega}_{j}^{L}-\Omega_{j}^{*})^{\top}\widehat{S}^{\mathcal{A}}(\widehat{\Omega}_{j}^{L}-\Omega_{j}^{*})\leqslant \frac{1}{2}(\Omega_{j}-\Omega_{j}^{*})^{\top}\widehat{S}^{\mathcal{A}}(\Omega_{j}-\Omega_{j}^{*})\\
			&\text{\quad}+(\Omega_j-\widehat{\Omega}_j^L)^{\top}(\widehat{S}^{\mathcal{A}}\Omega_{j}^{*}-(e_{j}+r_{j}(\widehat{\Delta}^{\mathcal{A}})))+\lambda_{\Omega}\|\Omega_{j}\|_{1}-\lambda_{\Omega}\|\widehat{\Omega}_{j}^{L}\|_{1}\\
			\Longrightarrow& \frac{1}{2} (\widehat{\Omega}_{j}^{L}-\Omega_{j}^{*})^{\top}\widehat{S}^{\mathcal{A}}(\widehat{\Omega}_{j}^{L}-\Omega_{j}^{*})\leqslant \frac{1}{2}(\Omega_{j}-\Omega_{j}^{*})^{\top}\widehat{S}^{\mathcal{A}}(\Omega_{j}-\Omega_{j}^{*})\\
			&\text{\quad}+|(\widehat{\Omega}_j^L-\Omega_j)^{\top}(\widehat{S}^{\mathcal{A}}\Omega_{j}^{*}-(e_{j}+r_{j}(\widehat{\Delta}^{\mathcal{A}})))|+\lambda_{\Omega}\|\Omega_{j}\|_{1}-\lambda_{\Omega}\|\widehat{\Omega}_{j}^{L}\|_{1}.
		\end{aligned}$$
		Consider the event $\mathcal{E}_{1}=\big\{\|\widehat{S}^{\mathcal{A}} \Omega_{j}^{*}-(e_{j}-r_{j}(\widehat{\Delta}^{\mathcal{A}}))\|_{\infty} \leqslant \lambda_{\Omega} / 2\big\}$. For this event, we have
		$$
		\begin{aligned}
			\frac{1}{2}(\widehat{\Omega}_{j}^{L}-\Omega_{j}^{*})^{\top}\widehat{S}^{\mathcal{A}}(\widehat{\Omega}_{j}^{L}-\Omega_{j}^{*}) \leqslant & \frac{1}{2}(\Omega_{j}-\Omega_{j}^{*})^{\top}\widehat{S}^{\mathcal{A}}(\Omega_{j}-\Omega_{j}^{*})+\frac{\lambda_{\Omega}}{2}\|\widehat{\Omega}_{j}^{L}-\Omega_{j}\|_{1}+\lambda_{\Omega}\|\Omega_{j}\|_{1}-\lambda_{\Omega}\|\widehat{\Omega}_{j}^{L}\|_{1} \\
			=&\frac{1}{2}(\Omega_{j}-\Omega_{j}^{*})^{\top}\widehat{S}^{\mathcal{A}}(\Omega_{j}-\Omega_{j}^{*})+\frac{\lambda_{\Omega}}{2}\|\widehat{\Omega}_{S_j,j}^{L}-\Omega_{S_j,j}\|_{1}+\frac{\lambda_{\Omega}}{2}\|\widehat{\Omega}_{S_j^c,j}^{L}\|_1\\
			&\text{\quad}+\lambda_{\Omega}\|\Omega_{S_j,j}\|_1-\lambda_{\Omega}\|\widehat{\Omega}_{S_j,j}^L\|_1-\lambda_{\Omega}\|\widehat{\Omega}_{S_j^c,j}^L\|_1\\
			\leqslant& \frac{1}{2}(\Omega_{j}-\Omega_{j}^{*})^{\top}\widehat{S}^{\mathcal{A}}(\Omega_{j}-\Omega_{j}^{*})+\frac{3 \lambda_{\Omega}}{2}\|\widehat{\Omega}_{S_{j},j}^{L}-\Omega_{S_{j},j}\|_{1}-\frac{\lambda_{\Omega}}{2}\|\widehat{\Omega}_{S_{j}^{c},j}^{L}-\Omega_{S_{j}^{c},j}\|_{1},
		\end{aligned}
		$$
		where we use the inequality $\|x\|_1-\|y\|_1 \leqslant \|x-y\|_1$ for vectors $x$, $y$ in the last line.
		The left hand side can be lower bounded by
		\begin{equation}\label{lb}
			\frac{1}{4}(\widehat{\Omega}_{j}^{L}-\Omega_{j})^{\top} \widehat{S}^{\mathcal{A}}(\widehat{\Omega}_{j}^{L}-\Omega_{j})-\frac{1}{2}(\Omega_{j}-\Omega_{j}^{*})^{\top} \widehat{S}^{\mathcal{A}}(\Omega_{j}-\Omega_{j}^{*}).
		\end{equation}
		Actually, due to positive semidefinite property of $\widehat{S}^{\mathcal{A}}$, we have
		$$\begin{aligned}
			&\text{\quad}\big[(\widehat{\Omega}_{j}^{L}-\Omega_j^*)-(\Omega_j^*-\Omega_j)\big]^{\top}\widehat{S}^{\mathcal{A}}\big[(\widehat{\Omega}_{j}^{L}-\Omega_j^*)-(\Omega_j^*-\Omega_j)\big]\\
			&=(\widehat{\Omega}_{j}^{L}-\Omega_j^*)^{\top}\widehat{S}^{\mathcal{A}}(\widehat{\Omega}_{j}^{L}-\Omega_j^*)+(\Omega_j^*-\Omega_j)^{\top}\widehat{S}^{\mathcal{A}}(\Omega_j^*-\Omega_j)-2(\widehat{\Omega}_{j}^{L}-\Omega_j^*)^{\top}\widehat{S}^{\mathcal{A}}(\Omega_j^*-\Omega_j)\\
			&\geqslant 0.
		\end{aligned}$$
		Hence, $2(\widehat{\Omega}_{j}^{L}-\Omega_j^*)^{\top}\widehat{S}^{\mathcal{A}}(\Omega_j^*-\Omega_j)\leqslant(\widehat{\Omega}_{j}^{L}-\Omega_j^*)^{\top}\widehat{S}^{\mathcal{A}}(\widehat{\Omega}_{j}^{L}-\Omega_j^*)+(\Omega_j^*-\Omega_j)^{\top}\widehat{S}^{\mathcal{A}}(\Omega_j^*-\Omega_j)$. Remark that if $\widehat{S}^{\mathcal{A}}$ is not positive semidefinite, we could make a projection of it into the space of semipositive definite matrices. On the other hand,
		$$\begin{aligned}
			&\text{\quad}(\widehat{\Omega}_{j}^{L}-\Omega_j)^{\top}\widehat{S}^{\mathcal{A}}(\widehat{\Omega}_{j}^{L}-\Omega_j)\\
			&=\big[(\widehat{\Omega}_{j}^{L}-\Omega_j^*)+(\Omega_j^*-\Omega_j)\big]^{\top}\widehat{S}^{\mathcal{A}}\big[(\widehat{\Omega}_{j}^{L}-\Omega_j^*)+(\Omega_j^*-\Omega_j)\big]\\
			&=(\widehat{\Omega}_{j}^{L}-\Omega_j^*)^{\top}\widehat{S}^{\mathcal{A}}(\widehat{\Omega}_{j}^{L}-\Omega_j^*)+(\Omega_j^*-\Omega_j)^{\top}\widehat{S}^{\mathcal{A}}(\Omega_j^*-\Omega_j)+2(\widehat{\Omega}_{j}^{L}-\Omega_j^*)^{\top}\widehat{S}^{\mathcal{A}}(\Omega_j^*-\Omega_j)\\
			&\leqslant 2(\widehat{\Omega}_{j}^{L}-\Omega_j^*)^{\top}\widehat{S}^{\mathcal{A}}(\widehat{\Omega}_{j}^{L}-\Omega_j^*)+2(\Omega_j^*-\Omega_j)^{\top}\widehat{S}^{\mathcal{A}}(\Omega_j^*-\Omega_j),
		\end{aligned}$$ which concludes (\ref{lb}).
		As a result,
		\begin{equation}\label{eqa.3}
			\begin{aligned}
				\frac{1}{4}(\widehat{\Omega}_{j}^{L}-\Omega_{j})^{\top}\widehat{S}^{\mathcal{A}}(\widehat{\Omega}_{j}^{L}-\Omega_{j}) \leqslant & (\Omega_{j}-\Omega_{j}^{*})^{\top} \widehat{S}^{\mathcal{A}}(\Omega_{j}-\Omega_{j}^{*}) \\
				&+\frac{3 \lambda_{\Omega}}{2}\|\widehat{\Omega}_{S_{j},j}^{L}-\Omega_{S_{j},j}\|_{1}-\frac{\lambda_{\Omega}}{2} \| \widehat{\Omega}_{S_{j}^{c},j}^{L}-\Omega_{S_{j}^{c},j} \|_{1}.
			\end{aligned}
		\end{equation}
		Let's consider the following two cases.\\	
		(i) If
		$$ \frac{3 \lambda_{\Omega}}{2} \| \widehat{\Omega}_{S_{j},j}^{L}-\Omega_{S_{j},j} \|_{1} \geqslant (\Omega_{j}-\Omega_{j}^{*})^{\top}\widehat{S}^{\mathcal{A}}(\Omega_{j}-\Omega_{j}^{*}), $$
		then inequality (\ref{eqa.3}) can be further relaxed into the following two forms.
		\begin{align}
			&\frac{1}{4}(\widehat{\Omega}_{j}^{L}-\Omega_{j})^{\top}\widehat{S}^{\mathcal{A}}(\widehat{\Omega}_{j}^{L}-\Omega_{j}) \leqslant  3\lambda_{\Omega}\|\widehat{\Omega}_{S_{j},j}^{L}-\Omega_{S_{j},j}\|_{1}-\frac{\lambda_{\Omega}}{2} \| \widehat{\Omega}_{S_{j}^{c},j}^{L}-\Omega_{S_{j}^{c},j} \|_{1} \label{casei1}\\
			&\frac{1}{4}(\widehat{\Omega}_{j}^{L}-\Omega_{j})^{\top}\widehat{S}^{\mathcal{A}}(\widehat{\Omega}_{j}^{L}-\Omega_{j}) \leqslant 3\lambda_{\Omega}\|\widehat{\Omega}_{S_{j},j}^{L}-\Omega_{S_{j},j}\|_{1}\label{casei2}
		\end{align}
		Using (\ref{casei1}) and taking advantage of $\widehat{S}^{\mathcal{A}}$'s positive semidefinite property, we have
		$$\|\widehat{\Omega}_{S_{j}^{c},j}^{L}-\Omega_{S_{j}^{c},j}\|_{1} \leqslant 6 \| \widehat{\Omega}_{S_{j},j}^{L}-\Omega_{S_{j},j} \|_{1}.$$
		In the event $\mathcal{E}_{2}=\bigg\{\inf\limits_{\|u_{S_{j}^{c}}\|_1 \leqslant 6\| u_{S_{j}} \|_{1} \neq 0} \frac{u^{\top} \widehat{S}^{\mathcal{A}} u}{\|u\|_{2}^{2}} \geqslant \phi_{0}>0\bigg\}$, we have
		$$\phi_0\|\widehat{\Omega}_{ j}^{L}-\Omega_{j}\|_{2}^{2} \leqslant(\widehat{\Omega}_{j}^{L}-\Omega_{j})^{\top}\widehat{S}^{\mathcal{A}}(\widehat{\Omega}_{j}^{L}-\Omega_{j}).$$
		Combining (\ref{casei2}), we obtain
		$$
		\begin{aligned}
			\frac{\phi_{0}}{4}\|\widehat{\Omega}_{ j}^{L}-\Omega_{j}\|_{2}^{2} \leqslant 3 \lambda_{\Omega}\|\widehat{\Omega}_{S_{j},j}^{L}-\Omega_{S_{j}, j}\|_{1} \leqslant 3 \sqrt{s_{j}} \lambda_{\Omega} \| \widehat{\Omega}_{S_{j},j}^{L}-\Omega_{S_{j},j} \|_{2}
			\leqslant 3 \sqrt{s_{j}} \lambda_{\Omega} \| \widehat{\Omega}_j^{L}-\Omega_j \|_{2},
		\end{aligned}
		$$
		which gives
		$$
		\|\widehat{\Omega}_{j}^{L}-\Omega_{j}\|_{2} \leqslant 12 \sqrt{s_{j}} \lambda_{\Omega} / \phi_{0}.
		$$
		(ii) If
		\begin{equation}\label{caseiicd}
			\begin{aligned}
				\frac{3 \lambda_{\Omega}}{2} \| \widehat{\Omega}_{S_{j}, j}^{L}-\Omega_{S_{j}, j} \|_{1} \leqslant (\Omega_{j}-\Omega_{j}^{*})^{\top}\widehat{S}^{\mathcal{A}}(\Omega_{j}-\Omega_{j}^{*}),
			\end{aligned}
		\end{equation}
		then inequality (\ref{eqa.3}) also can be further relaxed into the following two forms.
		\begin{align}
			&\frac{1}{4}(\widehat{\Omega}_{j}^{L}-\Omega_{j})^{\top}\widehat{S}^{\mathcal{A}}(\widehat{\Omega}_{j}^{L}-\Omega_{j}) \leqslant  2(\Omega_{j}-\Omega_{j}^{*})^{\top}\widehat{S}^{\mathcal{A}}(\Omega_{j}-\Omega_{j}^{*})-\frac{\lambda_{\Omega}}{2} \| \widehat{\Omega}_{S_{j}^{c} j}^{L}-\Omega_{S_{j}^{c}, j} \|_{1} \label{caseii1}\\
			&\frac{1}{4}(\widehat{\Omega}_{j}^{L}-\Omega_{j})^{\top}\widehat{S}^{\mathcal{A}}(\widehat{\Omega}_{j}^{L}-\Omega_{j}) \leqslant  2(\Omega_{j}-\Omega_{j}^{*})^{\top}\widehat{S}^{\mathcal{A}}(\Omega_{j}-\Omega_{j}^{*})\label{caseii2}
		\end{align}
		Again, using positive semidefinite property of $\widehat{S}^{\mathcal{A}}$ and (\ref{caseii1}), we have
		\begin{equation}\label{caseiiSc}
			\lambda_{\Omega} \| \widehat{\Omega}_{S_j^c,j}^{L}-\Omega_{S_j^c,j} \|_{1} \leqslant 4(\Omega_{j}-\Omega_{j}^{*})^{\top} \widehat{S}^{\mathcal{A}}(\Omega_{j}-\Omega_{j}^{*}).
		\end{equation}
		Combining (\ref{caseiicd}) and (\ref{caseiiSc}), we obtain $$\lambda_{\Omega} \| \widehat{\Omega}_{j}^{L}-\Omega_{j} \|_{1}=\lambda_{\Omega} \| \widehat{\Omega}_{S_j,j}^{L}-\Omega_{S_j,j} \|_{1}+\lambda_{\Omega} \| \widehat{\Omega}_{S_j^c,j}^{L}-\Omega_{S_j^c,j} \|_{1}\leqslant \frac{14}{3}(\Omega_{j}-\Omega_{j}^{*})^{\top}\widehat{S}^{\mathcal{A}}(\Omega_{j}-\Omega_{j}^{*}).
		$$
		Then we bound the term on the RHS
		$$
		\begin{aligned}
			&(\Omega_{j}-\Omega_{j}^{*})^{\top} \widehat{S}^{\mathcal{A}}(\Omega_{j}-\Omega_{j}^{*}) =(r_{j}(\widehat{\Delta}^{\mathcal{A}}-\Delta^{\mathcal{A}}))^{\top}\Omega^{\mathcal{A}} \widehat{S}^{\mathcal{A}} \Omega^{\mathcal{A}} r_{j}(\widehat{\Delta}^{\mathcal{A}}-\Delta^{\mathcal{A}}) \\
			&\leqslant (r_{j}(\widehat{\Delta}^{\mathcal{A}}-\Delta^{\mathcal{A}}))^{\top}\Omega^{\mathcal{A}} r_{j}(\widehat{\Delta}^{\mathcal{A}}-\Delta^{\mathcal{A}})+\|\Omega^{\mathcal{A}} (\widehat{S}^{\mathcal{A}}-\widehat{S}^{\mathcal{A}}) \Omega^{\mathcal{A}}\|_{\infty, \infty}\|r_{j}(\widehat{\Delta}^{\mathcal{A}}-\Delta^{\mathcal{A}})\|_1^2 \\
			&\leqslant ch\delta_n
		\end{aligned}
		$$
		with probability at least $1-\exp(c_1\log p)$.
		For any $\|u\|_{1} \leqslant \frac{14}{3}(\Omega_{j}-\Omega_{j}^{*})^{\top}\widehat{S}^{\mathcal{A}}(\Omega_{j}-\Omega_{j}^{*})/ \lambda_{\Omega}$, it holds that
		$$
		u^{\top} \widehat{S}^{\mathcal{A}} u \geqslant u^{\top} \Sigma^{\mathcal{A}} u -c_1\|\widehat{S}^{\mathcal{A}}-\Sigma^{\mathcal{A}}\|_{\max}\big((\Omega_{j}-\Omega_{j}^{*})^{\top}\widehat{S}^{\mathcal{A}}(\Omega_{j}-\Omega_{j}^{*})\big)^2.
		$$
		Therefore,
		$$
		\Lambda_{\min}(\Sigma^{A})\|\widehat{\Omega}_{j}^{L}-\Omega_{j}\|_{2}^{2} \leqslant c(\Omega_{j}-\Omega_{j}^{*})^{\top}\widehat{S}^{\mathcal{A}}(\Omega_{j}-\Omega_{j}^{*})\leqslant ch\delta_n
		$$ provided that $h\delta_n/\lambda_{\Omega}=O(1)$.
		To summarize, in event $\mathcal{E}=\{\mathcal{E}_1\cap\mathcal{E}_2\}$, we have
		$$
		\|\widehat{\Omega}_{j}^{L}-\Omega_{j}\|_{2}^{2} \leqslant c_1s \lambda_{\Omega}^{2}+c_2h\delta_n.
		$$
		It remains to lower bound the probability of the event $\mathcal{E}=\{\mathcal{E}_1\cap\mathcal{E}_2\}$. We have the following decomposition.
		$$
		\begin{aligned}
			&\widehat{S}^{\mathcal{A}} \Omega_{j}^{*}-(e_{j}+r_{j}(\widehat{\Delta}^{\mathcal{A}}))=\widehat{S}^{\mathcal{A}} \Omega_{j}+\widehat{S}^{\mathcal{A}} \Omega^{\mathcal{A}} r_{j}(\widehat{\Delta}^{\mathcal{A}}-\Delta^{\mathcal{A}})-e_{j}-r_{j}(\widehat{\Delta}^{\mathcal{A}}) \\
			&=\widehat{S}^{\mathcal{A}} \Omega_{j}+\widehat{S}^{\mathcal{A}} \Omega^{\mathcal{A}} r_{j}(\widehat{\Delta}^{\mathcal{A}}-\Delta^{\mathcal{A}})-e_{j}-r_{j}(\widehat{\Delta}^{\mathcal{A}})-r_j(\Delta^{\mathcal{A}})+r_j(\Delta^{\mathcal{A}}) \\
			&=\widehat{S}^{\mathcal{A}} \Omega_{j}-e_{j}-r_{j}(\Delta^{\mathcal{A}})+(\widehat{S}^{\mathcal{A}} \Omega^{\mathcal{A}}-I_{p}) r_{j}(\widehat{\Delta}^{\mathcal{A}}-\Delta^{\mathcal{A}}) \\
			&=(\widehat{S}^{\mathcal{A}}-\Sigma^{\mathcal{A}}) \Omega_{j}+(\widehat{S}^{\mathcal{A}} \Omega^{\mathcal{A}}-I_{p}) r_{j}(\widehat{\Delta}^{\mathcal{A}}-\Delta^{\mathcal{A}})
		\end{aligned}
		$$
		Using $\|AB\|_{\max}\leqslant\|A\|_{\max}\|B\|_{1}$, we have
		$$\begin{aligned}
			&\|(\widehat{S}^{\mathcal{A}}-\Sigma^{\mathcal{A}}) \Omega_{j}\|_{\max} \leqslant \|\widehat{S}^{\mathcal{A}}-\Sigma^{\mathcal{A}}\|_{\max}\|\Omega_j\|_1 \leqslant c_{1} \sqrt{\frac{\log p}{n_{\mathcal{A}}}} \\
			&\|(\widehat{S}^{\mathcal{A}} \Omega^{\mathcal{A}}-I_{p}) r_{j}(\widehat{\Delta}^{\mathcal{A}}-\Delta^{\mathcal{A}})\|_{\max} \leqslant \|\widehat{S}^{\mathcal{A}} \Omega^{\mathcal{A}}-I_{p}\|_{\max}\|r_{j}(\widehat{\Delta}^{\mathcal{A}}-\Delta^{\mathcal{A}})\|_{1} \\
			&\text{\qquad}\leqslant \|\widehat{S}^{\mathcal{A}}-\Sigma^{\mathcal{A}}\|_{\max}\| \Omega^{\mathcal{A}}\|_{1}\|r_{j}(\widehat{\Delta}^{\mathcal{A}}-\Delta^{\mathcal{A}})\|_{1} \leqslant c_{2}h\sqrt{\frac{\log p}{n_{\mathcal{A}}}}
		\end{aligned}
		$$
		for some constant $c_1$ and $c_2$ with probability at least $1-\exp(-c_3\log p)$.
		By $h \lesssim s \sqrt{\log p / n} \leqslant c$, it follows from the above results that $$\begin{aligned}
			\|\widehat{S}^{\mathcal{A}} \Omega_{j}^{*}-(e_{j}+r_{j}(\widehat{\Delta}^{\mathcal{A}}))\|_{\max} \leqslant & \|(\widehat{S}^{\mathcal{A}}-\Sigma^{\mathcal{A}}) \Omega_{j}\|_{\max}+\|(\widehat{S}^{\mathcal{A}} \Omega^{\mathcal{A}}-I_{p}) r_{j}(\widehat{\Delta}^{\mathcal{A}}-\Delta^{\mathcal{A}})\|_{\max} \\
			\leqslant &c_{4}\sqrt{\frac{\log p}{n_{\mathcal{A}}}}
		\end{aligned}$$
		with probability at least $1-\exp (-c_{3} \log p)$. Hence, $\mathbb{P}(\mathcal{E}_1)\geqslant1-\exp(-c_{3}\log p)$ for $\lambda_{\Omega}\geqslant2c_4\sqrt{\frac{\log p}{n_{\mathcal{A}}}}$. Using lemma \ref{REC}, it is easy to show that $\mathbb{P}(\mathcal{E}_2)\geqslant1-\exp(-c_{1}\log p)$ provided that $s^2\log p=o(n_{\mathcal{A}})$. This completes the proof of the first result.
		
		Then, we establish the convergence rate of $\widehat{\Omega}$ based on $\widehat{\Omega}^{L}$. According to (\ref{oplasso}), we have oracle inequality:
		$$\begin{aligned}
			\frac{1}{2}&(\widehat{\Omega}_{j}^{L})^{\top}\widehat{S}^{\mathcal{A}} \widehat{\Omega}_{j}^{L}-(\widehat{\Omega}_{j}^{L})^{\top}(e_{j}+r_{j}(\widehat{\Delta}^{\mathcal{A}}))+\lambda_{\Omega}\|\widehat{\Omega}_{j}^{L}\|_{1}\\
			&\leqslant \frac{1}{2}(\widehat{\Omega}_{j})^{\top}\widehat{S}^{\mathcal{A}} \widehat{\Omega}_{j}-\widehat{\Omega}_{j}^{\top}(e_{j}+r_{j}(\widehat{\Delta}^{\mathcal{A}}))+\lambda_{\Omega}\|\widehat{\Omega}_{j}\|_{1}.
		\end{aligned}$$
		As $\widehat{\Omega}_{j}^{L}$ is a feasible solution to (\ref{opS2p}), we have $\|\widehat{\Omega}_{j}\|_{1} \leqslant\|\widehat{\Omega}_{j}^{L}\|_{1}$. This leads to the following inequality
		$$\begin{aligned}
			\frac{1}{2}(\widehat{\Omega}_{j}^{L})^{\top}\widehat{S}^{\mathcal{A}}\widehat{\Omega}_{j}^{L}-(\widehat{\Omega}_{j}^{L})^{\top}(e_{j}+r_{j}(\widehat{\Delta}^{\mathcal{A}}))\leqslant \frac{1}{2}\widehat{\Omega}_{j}^{\top}\widehat{S}^{\mathcal{A}} \widehat{\Omega}_{j}-\widehat{\Omega}_{j}^{\top}(e_{j}+r_{j}(\widehat{\Delta}^{\mathcal{A}})),
		\end{aligned}$$
		and it can be transformed into the following form
		\begin{align}
			&\frac{1}{2}(\widehat{\Omega}_{j}^{L})^{\top}\widehat{S}^{\mathcal{A}} \widehat{\Omega}_{j}^{L}+\frac{1}{2}\widehat{\Omega}_{j}^{\top}\widehat{S}^{\mathcal{A}} \widehat{\Omega}_{j}-\widehat{\Omega}_j^{\top}\widehat{S}^{\mathcal{A}} \widehat{\Omega}_{j}^{L} \nonumber \\
			&\text{\quad}\leqslant \widehat{\Omega}_{j}^{\top}\widehat{S}^{\mathcal{A}} \widehat{\Omega}_{j}-\widehat{\Omega}_j^{\top}\widehat{S}^{\mathcal{A}} \widehat{\Omega}_{j}^{L}+(\widehat{\Omega}_{j}^L-\widehat{\Omega}_{j})^{\top}(e_{j}+r_{j}(\widehat{\Delta}^{\mathcal{A}})) \nonumber \\
			\Longrightarrow& \frac{1}{2} (\widehat{\Omega}_{j}^{L}-\widehat{\Omega}_{j})^{\top}\widehat{S}^{\mathcal{A}}(\widehat{\Omega}_{j}^{L}-\widehat{\Omega}_{j})\leqslant (\widehat{\Omega}_{j}-\widehat{\Omega}_{j}^L)^{\top}(\widehat{S}^{\mathcal{A}} \widehat{\Omega}_{j}-(e_{j}+r_{j}(\widehat{\Delta}^{\mathcal{A}}))) \nonumber \\
			\Longrightarrow& (\widehat{\Omega}_{j}-\widehat{\Omega}_{j}^{L})^{\top} \widehat{S}^{\mathcal{A}}(\widehat{\Omega}_{j}-\widehat{\Omega}_{j}^{L}) \leqslant 2 \lambda_{\Omega} \| \widehat{\Omega}_{j}-\widehat{\Omega}_{j}^{L} \|_{1}. \label{eqrec}
		\end{align}
		The inequality $\|\widehat{\Omega}_{j}\|_{1} \leqslant\|\widehat{\Omega}_{j}^{L}\|_{1}$ is equivalent to $$\|\widehat{\Omega}_{S_j,j}\|_{1}+\|\widehat{\Omega}_{S_j^c,j}\|_{1} \leqslant\|\widehat{\Omega}_{S_j,j}^{L}\|_{1}+\|\widehat{\Omega}_{S_j^c,j}^{L}\|_{1}$$ and from that we can obtain
		\begin{equation}\label{eqtl}
			\begin{aligned}
				\|\widehat{\Omega}_{S_{j}^{c},j}-\widehat{\Omega}_{S_{j}^{c},j}^{L}\|_{1} \leqslant& \|\widehat{\Omega}_{S_{j}^{c},j}\|_{1}+\|\widehat{\Omega}_{S_{j}^{c},j}^{L}\|_{1} \leqslant \|\widehat{\Omega}_{S_{j},j}^{L}\|_{1}-\|\widehat{\Omega}_{S_{j},j}\|_{1}+2\|\widehat{\Omega}_{S_{j}^{c},j}^{L}\|_{1}\\
				\leqslant& \|\widehat{\Omega}_{S_{j}}-\widehat{\Omega}_{S_{j},j}^{L}\|_{1}+2 \| \widehat{\Omega}_{S_{j}^{c},j}^{L} \|_{1}
			\end{aligned}
		\end{equation} where we use the inequality $\|x\|_1-\|y\|_1 \leqslant \|x-y\|_1$ again.
		Hence, we have $$\begin{aligned}
			\| \widehat{\Omega}_{j}-\widehat{\Omega}_{j}^{L} \|_{1} =& \|\widehat{\Omega}_{S_{j},j}-\widehat{\Omega}_{S_{j},j}^{L}\|_{1}+ \|\widehat{\Omega}_{S_{j}^{c},j}-\widehat{\Omega}_{S_{j}^{c},j}^{L} \|_{1} \\
			\leqslant& 2\|\widehat{\Omega}_{S_{j},j}-\widehat{\Omega}_{S_{j},j}^{L}\|_{1}+2 \| \widehat{\Omega}_{S_{j}^{c},j}^{L} \|_{1}
		\end{aligned}$$
		We can then separately discuss the two cases: $\|\widehat{\Omega}_{S_{j},j}-\widehat{\Omega}_{S_{j},j}^{L}\|_{1} \geqslant \|\widehat{\Omega}_{S_{j}^{c},j}^{L}\|_{1}$ and $\| \widehat{\Omega}_{S_{j},j}-\widehat{\Omega}_{S_{j},j}^{L}\|_{1} \leqslant  \| \widehat{\Omega}_{S_{j},j}^{L}\|_{1}.$\\
		(a)If $$\|\widehat{\Omega}_{S_{j},j}-\widehat{\Omega}_{S_{j},j}^{L}\|_{1} \geqslant \|\widehat{\Omega}_{S_{j}^{c},j}^{L}\|_{1},$$
		then according to (\ref{eqtl}), we have $$\|\widehat{\Omega}_{S_{j}^c,j}-\widehat{\Omega}_{S_{j}^c,j}^{L}\|_{1} \leqslant 3\|\widehat{\Omega}_{S_{j},j}-\widehat{\Omega}_{S_{j},j}^{L}\|_{1}$$ and $$\| \widehat{\Omega}_{j}-\widehat{\Omega}_{j}^{L} \|_{1} \leqslant 4\|\widehat{\Omega}_{S_{j},j}-\widehat{\Omega}_{S_{j},j}^{L}\|_{1}.$$
		Using lemma \ref{REC}, we have $$(\widehat{\Omega}_{j}-\widehat{\Omega}_{j}^{L})^{\top} \widehat{S}^{\mathcal{A}}(\widehat{\Omega}_{j}-\widehat{\Omega}_{j}^{L}) \geqslant (\widehat{\Omega}_{j}-\widehat{\Omega}_{j}^{L})^{\top} \Sigma^{\mathcal{A}}(\widehat{\Omega}_{j}-\widehat{\Omega}_{j}^{L}) -cs\sqrt{\frac{\log p}{n_{\mathcal{A}}}}\|\widehat{\Omega}_{j}-\widehat{\Omega}_{j}^{L}\|_2^2.$$ Invoking that under our assumptions, $s\sqrt{\frac{\log p}{n_{\mathcal{A}}}}=o(1)$, then $$c\| \widehat{\Omega}_{j}-\widehat{\Omega}_{j}^{L} \|_{2}^2\leqslant(\widehat{\Omega}_{j}-\widehat{\Omega}_{j}^{L})^{\top} \widehat{S}^{\mathcal{A}}(\widehat{\Omega}_{j}-\widehat{\Omega}_{j}^{L}).$$
		Combining (\ref{eqrec}), we have
		$$c\| \widehat{\Omega}_{j}-\widehat{\Omega}_{j}^{L} \|_{2}^2 \leqslant 8\lambda_{\Omega}\|\widehat{\Omega}_{S_{j},j}-\widehat{\Omega}_{S_{j},j}^{L}\|_{1}\leqslant 8 \sqrt{s}\lambda_{\Omega}\| \widehat{\Omega}_{j}-\widehat{\Omega}_{j}^{L} \|_{2},$$
		which gives $$\| \widehat{\Omega}_{j}-\widehat{\Omega}_{j}^{L} \|_{2}^2\leqslant c_1 s\lambda_{\Omega}^2.$$
		(b)If $$\|\widehat{\Omega}_{S_{j},j}-\widehat{\Omega}_{S_{j},j}^{L}\|_{1} \leqslant \|\widehat{\Omega}_{S_{j}^{c},j}^{L}\|_{1},$$
		then according to (\ref{eqtl}), we have $$\|\widehat{\Omega}_{S_{j}^c,j}-\widehat{\Omega}_{S_{j}^c,j}^{L}\|_{1} \leqslant 3\|\widehat{\Omega}_{S_{j}^{c},j}^{L}\|_{1}$$ and $$\| \widehat{\Omega}_{j}-\widehat{\Omega}_{j}^{L} \|_{1} \leqslant 4 \| \widehat{\Omega}_{S_{j}^{c},j}^{L} \|_{1}.$$
		If case (i) discussed above, we have
		$$
		\|\widehat{\Omega}_{S_{j}^{c},j}^{L}\|_{1}=\|\widehat{\Omega}_{S_{j}^{c},j}^{L}-\Omega_{S_{j}^{c},j}\|_{1} \leqslant 6\|\widehat{\Omega}_{S_{j},j}^{L}-\Omega_{S_{j},j}\|_{1} \leqslant c_{1} s \lambda_{\Omega}.
		$$
		In case (ii) discussed above, we have
		$$
		\lambda_{\Omega}\|\widehat{\Omega}_{S_{j}^{c},j}^{L}\|_{1} \leqslant 3(\Omega_{j}-\Omega_{j}^{*})^{\top} \widehat{S}^{\mathcal{A}}(\Omega_{j}-\Omega_{j}^{*}) \leqslant c_{2} h\delta_n.
		$$
		Therefore, combining (\ref{eqrec}), we could obtain $$(\widehat{\Omega}_{j}-\widehat{\Omega}_{j}^{L})^{\top} \widehat{S}^{\mathcal{A}}(\widehat{\Omega}_{j}-\widehat{\Omega}_{j}^{L})\leqslant c_1s\lambda_{\Omega}^2+c_2h\delta_n.$$
		Notice that $$(\widehat{\Omega}_{j}-\widehat{\Omega}_{j}^{L})^{\top} \Sigma^{\mathcal{A}}(\widehat{\Omega}_{j}-\widehat{\Omega}_{j}^{L})-\|\widehat{S}^{\mathcal{A}}-\Sigma^{\mathcal{A}}\|_{\max}\|\widehat{\Omega}_{j}-\widehat{\Omega}_{j}^{L}\|_1^2\leqslant(\widehat{\Omega}_{j}-\widehat{\Omega}_{j}^{L})^{\top} \widehat{S}^{\mathcal{A}}(\widehat{\Omega}_{j}-\widehat{\Omega}_{j}^{L})$$
		and $$\begin{aligned}
			\|\widehat{S}^{\mathcal{A}}-\Sigma^{\mathcal{A}}\|_{\max}\|\widehat{\Omega}_{j}-\widehat{\Omega}_{j}^{L}\|_1^2 \leqslant & c_3\sqrt{\frac{\log p}{n_{\mathcal{A}}}}(s\lambda_{\Omega}+h\delta_n/\lambda_{\Omega})^2\\
			\leqslant & c_4(s\lambda_{\Omega}^2+h\delta_n),
		\end{aligned}$$
		we have $$\| \widehat{\Omega}_{j}-\widehat{\Omega}_{j}^{L} \|_{2}^2 \leqslant c_1s\lambda_{\Omega}^2+c_2h\delta_n$$
		Using the inequation $\| \widehat{\Omega}_{j}-\Omega_j \|_{2}^2\leqslant 2\| \widehat{\Omega}_{j}-\widehat{\Omega}_{j}^{L} \|_{2}^2+2\| \widehat{\Omega}_{j}^{L}-\Omega_j \|_{2}^2$, we can easily obtain the desired results.
	\end{proof}

\subsection{Proof of Theorem \ref{t2}}	
\begin{proof}
Theorem \ref{t2} can be easily deduced from the results from Lemma \ref{CgvCL}.
\end{proof}
	\subsection{Proof of useful lemmas}
	Proof of Lemma \ref{CvgS}  It is adapted from Theorem 4.2 in \cite{liu2012high}, so we omit the proof here.\\
	Proof of Lemma \ref{CvgSA} Using Theorem 4.2 in \cite{liu2012high}, we have $$\mathbb{P}\bigg(\|\widehat{S}^{(k)}-\Sigma^{(k)}\|_{\max}\geqslant t\bigg)\leqslant p^2\exp(-\frac{n_kt^2}{2\pi^2}).$$
	Then
	$$\begin{aligned}
		\mathbb{P}\bigg(\|\widehat{S}^{\mathcal{A}}-\Sigma^{\mathcal{A}}\|_{\max}\geqslant t\bigg) &= \mathbb{P}\bigg(\|\sum\limits_{k\in\mathcal{A}}\alpha_k(\widehat{S}^{(k)}-\Sigma^{(k)})\|_{\max}\geqslant t\bigg) \\
		& \leqslant \mathbb{P}\bigg(\sum\limits_{k\in\mathcal{A}}\alpha_k\|\widehat{S}^{(k)}-\Sigma^{(k)}\|_{\max}\geqslant t\bigg) \\
		& \leqslant \sum\limits_{k\in\mathcal{A}}\mathbb{P}\bigg(\|\widehat{S}^{(k)}-\Sigma^{(k)}\|_{\max}\geqslant \frac{t}{|\mathcal{A}|\alpha_k}\bigg) \\
		&\leqslant p^2\sum\limits_{k\in\mathcal{A}}\exp\{-\frac{1}{2\pi^2}n_k\frac{t^2n_{\mathcal{A}}^2}{|\mathcal{A}|^2n_k^2}\} \\
		&\leqslant p^2\exp\{-cn_{\mathcal{A}}t^2\},
	\end{aligned}$$ which concludes the lemma.\\
	Proof of Lemma \ref{REC}. First, we have
	$$\begin{aligned}
		u^{\top}\Sigma u - u^{\top}\widehat{S}u =& u^{\top}(\Sigma-\widehat{S})u \\
		\leqslant& \|\Sigma-\widehat{S}\|_{\max}\|u\|_1^2 \\
		\leqslant& \tilde{\lambda}\|u\|_1^2.
	\end{aligned}$$
	Since $$\begin{aligned}
		\|u\|_1=&\|u_S\|_1+\|u_{S^c}\|_1\leqslant(1+\alpha)\|u_S\|_1\\
		\leqslant&\sqrt{s}(1+\alpha)\|u_S\|_2\leqslant\sqrt{s}(1+\alpha)\|u\|_2,
	\end{aligned}$$
	we can obtain $u^{\top}\Sigma u - u^{\top}\widehat{S}u\leqslant \tilde{\lambda}s(1+\alpha)^2\|u\|^2_2$.\\
	Proof of Lemma \ref{Cvgrj}. Using $\|AB\|_{\max}\leqslant\|A\|_{\max}\|B\|_{1}$, we have $$\|(\widehat{S}-\Sigma) \Delta^{\mathcal{A}}\|_{\max}\leqslant\|\widehat{S}-\Sigma\|_{\max}\|\Delta^{\mathcal{A}}\|_{1}\leqslant c_1 \sqrt{\frac{\log p}{n}}.$$
	The triangular inequality $\|A+B\|_{\max}\leqslant\|A\|_{\max}+\|B\|_{\max}$ entails that $$
	\begin{aligned}
		\|\Sigma \Delta^{\mathcal{A}}-(\widehat{S}^{\mathcal{A}}-\widehat{S})\|_{\max} =& \|(\Sigma^{\mathcal{A}}-\Sigma)-(\widehat{S}^{\mathcal{A}}-\widehat{S})\|_{\max}\\
		\leqslant& \|\Sigma^{\mathcal{A}}-\widehat{S}^{\mathcal{A}}\|_{\max}+\|\Sigma-\widehat{S}\|_{\max} \\
		\leqslant& c_2\sqrt{\frac{\log p}{n}},
	\end{aligned}
	$$
	and it follows from the above results that
	$$
	\begin{aligned}
		\|\widehat{S} \Delta^{\mathcal{A}}-(\widehat{S}^{\mathcal{A}}-\widehat{S})\|_{\max}
		\leqslant &\|(\widehat{S}-\Sigma) \Delta^{\mathcal{A}}\|_{\max}+\|\Sigma \Delta^{\mathcal{A}}-(\widehat{S}^{\mathcal{A}}-\widehat{S})\|_{\max} \\
		\leqslant& c_3\sqrt{\frac{\log p}{n}}
	\end{aligned}
	$$
	for some constant $c_1$ and $c_2$ with probability at least $1-\exp(-c_4\log p)$. This shows that for $\lambda_{\Delta} \geqslant c_{1} \sqrt{\log p / n}$, $\Delta^{\mathcal{A}}$ is a feasible solution to $(\ref{opS1})$. Hence,
	$$
	\|\widehat{\Delta}_{j}^{(0)}\|_{1} \leqslant\|\Delta_{j}^{\mathcal{A}}\|_{1} \leqslant h,
	$$
	$$
	\mathbb{P}\bigg(\max _{j}\|\widehat{\Delta}_{j}^{(0)}-\Delta_{j}^{\mathcal{A}}\|_{1} \leqslant 2 h\bigg) \geqslant 1-\exp (-c_{4} \log p).
	$$
	Next we will show that $\Delta^{\mathcal{A}}$ is in the feasible set of (\ref{opS1db}) with probability at least $1-\exp (-c_{1} \log p)$. Define $
	\widehat{\Delta}^{(db)}=\widehat{\Delta}^{(0)}+\widehat{\Omega}^{(\mathrm{CL})}(\widehat{S}^{\mathcal{A}}-\widehat{S}-\widehat{S} \widehat{\Delta}^{(0)})
	$, this is essentially a debiased estimator of $\Delta^{\mathcal{A}}$. We have the following decomposition:
	$$\begin{aligned}
		\widehat{\Delta}^{(db)}-\Delta^{\mathcal{A}}=&\widehat{\Delta}^{(0 )}-\Delta^{\mathcal{A}}+\widehat{\Omega}^{(\mathrm{CL})}(\widehat{S}^{\mathcal{A}}-\widehat{S}-\widehat{S} \Delta^{\mathcal{A}})-\widehat{\Omega}^{(\mathrm{CL})} \widehat{S}(\widehat{\Delta}^{(0)}-\Delta^{\mathcal{A}})\\
		=&(I_{p}-\widehat{\Omega}^{(\mathrm{CL})} \widehat{S})(\widehat{\Delta}^{(0)}-\Delta^{\mathcal{A}})+\widehat{\Omega}^{(\mathrm{CL})}(\widehat{S}^{\mathcal{A}}-\Sigma^{\mathcal{A}}-(\widehat{S}-\Sigma)(I_{p}+\Delta^{\mathcal{A}}))\\
		=&\underbrace{(I_{p}-\widehat{\Omega}^{(\mathrm{CL})} \widehat{S})(\widehat{\Delta}^{(0)}-\Delta^{\mathcal{A}})}_{rem_{1}}+\underbrace{\Omega(\widehat{S}^{A}-\Sigma^{\mathcal{A}}-(\widehat{S}-\Sigma)(I_{p}+\Delta^{\mathcal{A}}))}_{rem_{2}}\\
		&+\underbrace{(\widehat{\Omega}^{(\mathrm{CL})}-\Omega)(\widehat{S}^{\mathcal{A}}-\Sigma^{\mathcal{A}}-(\widehat{S}-\Sigma)(I_{p}+\Delta^{\mathcal{A}}))}_{rem_{3}}.
	\end{aligned}$$
	Again, using $\|AB\|_{\max}\leqslant\|A\|_{\max}\|B\|_{1}$ and triangular inequality $\|A+B\|_{\max}\leqslant\|A\|_{\max}+\|B\|_{\max}$, we have
	$$
	\begin{aligned}
		&\|rem_{1}\|_{\max} \leqslant\|I_{p}-\widehat{\Omega}^{(\mathrm{CL})} \widehat{S}\|_{\max}\|\widehat{\Delta}^{(0)}-\Delta^{\mathcal{A}} \|_{1} \leqslant 2h \lambda_{\mathrm{CL}} \leqslant c_1 \frac{s \log p}{n}\\
		&\|\widehat{S}^{A}-\Sigma^{\mathcal{A}}-(\widehat{S}-\Sigma)(I_{p}+\Delta^{\mathcal{A}})\|_{\max} \leqslant \|\widehat{S}^{A}-\Sigma^{\mathcal{A}}\|_{\max}+\|\widehat{S}-\Sigma\|_{\max}\|I_{p}+\Delta^{\mathcal{A}}\|_{1}\leqslant c_2\sqrt{\frac{\log p}{n}}\\
		&\|rem_{2}\|_{\max} \leqslant \|\Omega\|_{1}\|\widehat{S}^{A}-\Sigma^{\mathcal{A}}-(\widehat{S}-\Sigma)(I_{p}+\Delta^{\mathcal{A}})\|_{\max} \leqslant c_3 \sqrt{\frac{\log p}{n}} \\
		&\|rem_{3}\|_{\max} \leqslant\|\widehat{\Omega}^{(\text{CL})}-\Omega\|_{1}\|\widehat{S}^{A}-\Sigma^{\mathcal{A}}-(\widehat{S}-\Sigma)\left(I_{p}+\Delta^{\mathcal{A}}\right) \|_{\max} \leqslant c_4 \frac{s \log p}{n}
	\end{aligned}
	$$
	with probability at least $1-\exp(-c_5\log p)$.
	By $h \lesssim s \sqrt{\log p / n} \leqslant c$, it follows from the above results that
	$$\begin{aligned}
		\|\widehat{\Delta}^{(db)}-\Delta^{\mathcal{A}}\|_{\max} \leqslant& \|rem_1\|_{\max}+\|rem_2\|_{\max}+\|rem_3\|_{\max} \\
		\leqslant &C \sqrt{\frac{\log p}{n}}
	\end{aligned}
	$$ for a large enough constant C with probability at least $1-\exp (-c_{5} \log p)$.
	According to the triangle inequality $$\|r_{j}(\widehat{\Delta}^{\mathcal{A}}-\Delta^{\mathcal{A}})\|_{\infty}\leqslant \|r_{j}(\widehat{\Delta}^{\mathcal{A}}-\Delta^{(db)})\|_{\infty}+\|r_{j}(\widehat{\Delta}^{(db)}-\Delta^{\mathcal{A}})\|_{\infty}\leqslant c\sqrt{\frac{\log p}{n}},$$ it is easy to see that $\Delta^{\mathcal{A}}$ is a feasible solution to (\ref{opS1db}) with probability at least $1-\exp (-c_{1} \log p)$.
	Using the fact that
	$$
	\|r_{j}(\widehat{\Delta}^{\mathcal{A}})\|_{1} \leqslant\|r_{j}(\Delta^{\mathcal{A}})\|_{1} \leqslant h
	$$
	we have $$
	\max_{j \leqslant p}\|r_{j}(\widehat{\Delta}^{\mathcal{A}}-\Delta^{\mathcal{A}})\|_{2}^{2} \leqslant \max_{j \leqslant p}\|r_{j}(\widehat{\Delta}^{\mathcal{A}}-\Delta^{\mathcal{A}})\|_{1} \|r_{j}(\widehat{\Delta}^{\mathcal{A}}-\Delta^{\mathcal{A}}) \|_{\infty} \leqslant c_1 h \sqrt{\frac{\log p}{n}}
	$$
	and $$ \max_{j \leqslant p}\|r_{j}(\widehat{\Delta}^{\mathcal{A}}-\Delta^{\mathcal{A}})\|_{2}^{2} \leqslant \max_{j \leqslant p}\|r_{j}(\widehat{\Delta}^{\mathcal{A}}-\Delta^{\mathcal{A}})\|_{1}^2  \leqslant c_2 h^2.$$
	Hence $$ \max_{j \leqslant p}\|r_{j}(\widehat{\Delta}^{\mathcal{A}}-\Delta^{\mathcal{A}})\|_{2}^{2} \leqslant C h\delta_{n}$$
	for $\delta_{n}=\sqrt{\frac{\log p}{n}} \wedge h$ with probability at least $1-\exp (-c_{1} \log p)$.\\
	Proof of Lemma \ref{CgvCL}. According to Lemma \ref{CvgS} and Theorem 9.6's proof process in \cite{fan2020statistical}, for $\lambda_{\mathrm{CL}} \geqslant c \sqrt{\log p / n}$ with large enough constant $c$, we have
	$$\|\widehat{\Omega}_{j}^{(\mathrm{CL})}-\Omega_{j}\|_1 \leqslant c_1 s\sqrt{\frac{\log p}{n}}$$ with probability at least $1-\exp\{-c_2\log p\}$ and $\Omega$ is in the feasible set of problem (\ref{opcl}). By Lemma 1 in \cite{cai2011constrained}, problem (\ref{opcl}) can be further casted as $p$ separate minimization problems and for $1 \leqslant j \leqslant p$,
	$$
	\begin{aligned}
		\widehat{\Omega}_{j}^{(\mathrm{CL})} &=\underset{\omega}{\arg \min }\|\omega\|_{1} \\
		& \text { subject to }\|\widehat{S} \omega-e_{j}\|_{\infty} \leqslant \lambda_{\mathrm{CL}}
	\end{aligned}
	$$Hence,
	\begin{equation}\label{eqcl}
		\begin{aligned}
			(\widehat{\Omega}_{j}^{(\mathrm{CL})}-\Omega_{j})^{\top} \widehat{S}(\widehat{\Omega}_{j}^{(\mathrm{CL})}-\Omega_{j}) \leqslant & \|\widehat{S}(\widehat{\Omega}_{j}^{(\mathrm{CL})}-\Omega_{j})\|_\infty\|\widehat{\Omega}_{j}^{(\mathrm{CL})}-\Omega_{j}\|_1 \\
			\leqslant & \big(\|\widehat{S}\widehat{\Omega}_{j}^{(\mathrm{CL})}-e_j\|_\infty+\|\widehat{S}\Omega_{j}-e_j\|_\infty\big)\|\widehat{\Omega}_{j}^{(\mathrm{CL})}-\Omega_{j}\|_1\\
			\leqslant &2 \lambda_{\mathrm{CL}} \| \widehat{\Omega}_{j}^{(\mathrm{CL})}-\Omega_{j} \|_{1}
		\end{aligned}
	\end{equation}
	Next we show that $\widehat{\Omega}_{j}^{(\mathrm{CL})}-\Omega_{j}\in \mathcal{C}(S_j,1)$, so we could use Lemma \ref{REC} to bound $\| \widehat{\Omega}_{j}^{(\mathrm{CL})}-\Omega_{j} \|_{2}^2$ by $(\widehat{\Omega}_{j}^{(\mathrm{CL})}-\Omega_{j})^{\top} \widehat{S}(\widehat{\Omega}_{j}^{(\mathrm{CL})}-\Omega_{j})$.
	Notice that $\|\widehat{\Omega}_{j}^{(\mathrm{CL})}\|_{1} \leqslant\|\Omega_{j}\|_{1}$ and
	$$\begin{aligned}
		\|\widehat{\Omega}^{\text{(CL)}}_j\|_1 =& \|\widehat{\Omega}^{\text{(CL)}}_j-\Omega_j+\Omega_j\|_1 = \|\widehat{\Omega}^{\text{(CL)}}_{S_j,j}-\Omega_{S_j,j}+\Omega_{S_j,j}\|_1+\|\widehat{\Omega}^{\text{(CL)}}_{S_j^c,j}-\Omega_{S_j,j}+\Omega_{S_j^c,j}\|_1\\
		\geqslant &\|\Omega_j\|_1-\|\widehat{\Omega}^{\text{(CL)}}_{S_j,j}-\Omega_{S_j,j}\|_1+\|\widehat{\Omega}^{\text{(CL)}}_{S_j^c,j}-\Omega_{S_j,j}\|_1,
	\end{aligned}$$
	so we have $\|\widehat{\Omega}^{\text{(CL)}}_{S_j^c,j}-\Omega_{S_j,j}\|_1 \leqslant \|\widehat{\Omega}^{\text{(CL)}}_{S_j,j}-\Omega_{S_j,j}\|_1$. Using lemma \ref{REC}, we have $$(\widehat{\Omega}_{j}^{(\mathrm{CL})}-\Omega_j)^{\top}\widehat{S}(\widehat{\Omega}_{j}^{(\mathrm{CL})}-\Omega_j) \geqslant (\widehat{\Omega}_{j}^{(\mathrm{CL})}-\Omega_j)^{\top}\Sigma (\widehat{\Omega}_{j}^{(\mathrm{CL})}-\Omega_j) -s\sqrt{\frac{\log p}{n}}\|(\widehat{\Omega}_{j}^{(\mathrm{CL})}-\Omega_j)\|_2^2.$$ Invoking that under our assumptions, $s\sqrt{\frac{\log p}{n}}=o(1)$, then $c\|(\widehat{\Omega}_{j}^{(\mathrm{CL})}-\Omega_j)\|_2^2 \leqslant (\widehat{\Omega}_{j}^{(\mathrm{CL})}-\Omega_j)^{\top}\widehat{S}(\widehat{\Omega}_{j}^{(\mathrm{CL})}-\Omega_j)$. Together with (\ref{eqcl}), we arrive at the desired results.
\end{appendices}

\bibliographystyle{model2-names}

\bibliography{ref}

\end{document}